\documentclass[traditabstract]{aa} 
\usepackage[english]{babel}
\usepackage{graphicx}
\usepackage{txfonts}
\usepackage{wasysym}
\newcommand{\gtap}{\mathrel{\hbox{\rlap{\lower.55ex \hbox {$\sim$}}
                   \kern-.3em \raise.4ex \hbox{$>$}}}}
\newcommand{\ltap}{\mathrel{\hbox{\rlap{\lower.55ex \hbox {$\sim$}}
                   \kern-.3em \raise.4ex \hbox{$<$}}}}
\newcommand{\psrbad}{PSR\,B1508+55}
\begin{document}
\selectlanguage{english}
  \title{The observed velocity distribution of young pulsars}

  \author{Frank Verbunt \and Andrei Igoshev \and Eric Cator}

  \institute{Institute of Mathematics, Astrophysics and Particle Physics, Radboud University Nijmegen, PO Box 9010,
    6500 GL Nijmegen, The Netherlands; 
      F.Verbunt@astro.ru.nl, \email{A.Igoshev@astro.ru.nl, E.Cator@science.ru.nl}
}

  \date{submitted to A\&A on 6 July 2017}

  \abstract{We argue that comparison with observations of theoretical
    models for the velocity distribution of pulsars must be done
    directly with the observed quantities, i.e.\ parallax and the two
    components of proper motion. We develop a formalism to do so, and
    apply it to pulsars with accurate VLBI measurements. We find that
    a distribution with two maxwellians improves significantly on
    a single maxwellian. The `mixed' model takes into account that pulsars move away
    from their place of birth, a narrow region around the galactic
    plane.  The best model has 42\%\ of the pulsars in a maxwellian
    with average velocity $\sigma\sqrt{8/\pi}=120$\,km/s, and 
    58\%\ in a maxwellian with average velocity 540\,km/s. About 5\%\
    of the pulsars has a velocity at birth less than 60\,km/s.
    For the youngest pulsars ($\tau_c<10$\,Myr), these numbers are 
    32\%\ with 130\,km/s, 68\%\ with 520\,km/s, and 3\%, with
    appreciable uncertainties. 
    }


 \maketitle

\section{Introduction}

The study of the velocities of pulsars is interesting on its own
account, as a pointer to the formation process of a neutron star,
but also has ramifications beyond this. In particular, some
neutron stars are found in binaries and in globular clusters, as accreting X-ray 
sources or as pulsars. These neutron stars were born with velocities
less than the escape velocity from the binary or from the cluster.

Neutron stars that have the same velocities as their progenitors,
move with the rotation of the galaxy, with small velocities
with respect to the local standard of rest (LSR), unless their
progenitor is a member of a close binary or a runaway star.
To investigate the velocities that neutron star acquires
at birth in addition to the progenitor velocity, one therefore
investigates their velocity $v$ with respect to the LSR.

This investigation is complicated for pulsars with large velocities as
these are affected by an acceleration in the galactic potential that
varies between their place of birth and their current location, and
because their current LSR differs from the LSR at their place of
birth. Thus the current $v$ of a pulsar differs from the $v$
at birth.  If the age and full space velocities were
known, we could solve this complication by integrating the pulsar
orbit back in time, but proper motion studies only provide 2 of the 3
velocity components, and ages of pulsars are usually uncertain.  By
limiting the study to young pulsars, one may reduce the effect of
these complications.
As well described by Brisken et al.\ (2003a, in particular Sect.5.1)\nocite{bfg+03},
correlations between spin-axis and velocity,  between 
luminosity and velocity, and/or between velocity and distance to the Galactic
Plane, among others, introduce selection effects in the observations.
Such selection effects can only be corrected for in a full population study.
Even so,  determining the observed $v$ distribution is a useful
step toward a full population study, and various efforts have been
published (see Table\,\ref{t:previous}).

Arzoumanian et al.\ (2002) \nocite{arzoumanian2002}
compare synthesized model populations 
with the observed periods, period derivatives, dispersion measures, 
fluxes, and the absolute values of galactic latitudes and of proper motions.
They conclude that the velocity distribution of pulsars is bimodal,
with a low-velocity and a high-velocity component. 

Brisken et al.\ (2003a) \nocite{bfg+03} investigate the velocity
component $v_l$ in the direction of galactic longitude. Their study is
based on interferometric proper motion measurements (mostly their
own).  For each pulsar, they compute a probability distribution $P(D)$
for the distance $D$ (based on the parallax or on the dispersion
measure DM, allowing for the limited accuracy in converting DM to
$D$) and combine this with the probability function $P(\mu_l)$ for the
proper motion $\mu_l$ (allowing for measurement uncertainty) to
compute the probability distribution $P(v_l)$.  The set of $P(v_l)$ is
fitted with a model in which this distribution is described by two
zero-centred Gaussian distributions, representing a slow and a fast
component.  

Hobbs et al.\ (2005) \nocite{hobbs2005} construct velocity
distributions $P(v_\mathrm{1D})$ where $v_\mathrm{1D}$ is either $v_l$
or $v_b$ and $P(v_\mathrm{2D})$ where
$v_\mathrm{2D}\equiv\sqrt{{v_l}^2+{v_b}^2}$ for a larger sample of
pulsars, including measurements based on timing. $v_b$ is the velocity
component in the direction of latitude. Hobbs et al.\ assume that these observed
$v_\mathrm{1D}$ and $v_\mathrm{2D}$ distributions are projections of
an isotropic velocity distribution $P(v)$, and then reconstruct $P(v)$
by using a {\tt clean} algorithm to deconvolve $P(v_\mathrm{1D})$ and
$P(v_\mathrm{2D})$.  The advantage of this method is that it is
non-parametric, i.e.\ it does not assume a prescribed form for $P(v)$.
The reconstructed form turns out to be well described by a
maxwellian, with $\sigma=265$\,km/s.

Faucher-Gigu\`ere \&\ Kaspi (2006) \nocite{faucher2006} extend
the method of Brisken et al.\ (2003a) in two ways. First they consider
a variety of models for the distribution of $v_l$, and second they extend the
maximum-likelihood model with a Bayesian analysis of
probability ratios for the comparison of different models.

Whereas these studies agree that the space (i.e.\ 3-D) velocities of
neutron stars are high, averaging as much as 450\,km/s, they differ on
the fraction of low-velocity neutron stars. Hobbs et al.\ (2005) argue
that the low-velocity tail of the pulsar velocity distribution is due
to projection effects, and that very few pulsars have space velocities
below 60 km/s. (For a maxwellian with $\sigma=265$\,km/s the fraction
is 0.003.)  In the acceptable models discussed by Faucher-Gigu\`ere
\&\ Kaspi (2006) the derived fraction of pulsars with space velocities
less than 60 km/s varies from 0.012 (for a two-component Gaussian) to 0.135
(for the Paczy\'nski distribution).

One reason for us to make a new study of the pulsar velocities is to
resolve the differences between the predicted numbers of low-velocity
pulsars in these recent studies. We note that among nine very accurate
pulsar velocities $v_\perp$ ($=\sqrt{ {v_l}^2+{v_b}^2}$) listed by Brisken et al.\ (2002,
Table\,5), two are smaller than 40\,km/s.  The probability of finding
two such low-$v_\perp$ pulsars in a sample of nine is 0.004 for an
isotropic maxwellian with $\sigma=265$\,km/s.  This suggests that the
pulsar velocities may be overestimated by Hobbs et al.\ (2005).

A second reason for a new study is the development by Verbiest et
al. (2012), of a Bayesian method to combine different distance
indicators into a single probability distribution $P(D)$ for each
pulsar. The main distance indicator is the parallax, where the
Lutz-Kelker (1973) \nocite{lutzkelker73} effect is taken into account,
with the galactic pulsar distribution as a prior.  For the study of
pulsar velocities we correct some errors in the equations given by
Verbiest et al.\ (2012) for use of the parallax (see Igoshev et al.\ 2016
\nocite{igoshev2016} and Bailer-Jones 2015 \nocite{bailer2015}), and add the
measurements of the proper motions.

The third and final reason for our new study of pulsar
velocities is the increased number of accurately measured proper motions and parallaxes
(see Table\,\ref{t:sources}).

In Section\,\ref{s:data} we describe the master list of observed
proper motions that we use in our study.  We describe the ingredients
of the likelihood function for pulsars and their use in determining
the parameters of the velocity distribution in Sect.\,\ref{s:ingredients},
and apply these to various models: a single isotropic maxwellian in
Sect.\,\ref{s:maxwell}, the sum of two isotropic maxwellians in
Sect.\,\ref{s:two_maxw},  and a mixture of one or two isotropic and
semi-isotropic maxwellians in Sect.\,\ref{s:mixed}.
(In the semi-isotropic maxwellian distribution velocities towards the galactic plane 
are excluded, as explained in Sect.\,\ref{s:semiiso}.)

Before we proceed, we describe the notation we use:
we differentiate between the actual (and generally
unknown) properties of the pulsar, and the measured (or nominal)
values, by indicating the latter with a prime ($'$).
The actual proper motion is the sum of three components:
one due to the peculiar velocity of the pulsar,  one due to
the difference between the local standards of rest of the pulsar and
of the Sun, and one due to the peculiar motion of the Sun
(Eqs.\,\ref{e:mucor} - \ref{e:theta}).
The measured parallax and proper motion differs from the actual values 
due to measurement errors (Eqs.\,\ref{e:erpar} - \ref{e:ermud}), and
may be skewed due non-uniform distributions of positions and velocities
(Fig.\,\ref{f:examples}).
For convenience, our notation is summarized in Table\ref{t:notation}.

\begin{table}
\caption{Notation used in this paper \label{t:notation}}
\centerline{
\begin{tabular}{lccc}
\multicolumn{4}{c}{\bf actual (unknown) values}\\ 
\hline
parallax, distance & $\varpi$ & $D=1/\varpi$ \\
& & equatorial & galactic \\
peculiar velocity & $v$ &  $v_\alpha,v_\delta,v_r$ &
  $v_l,v_b,v_r$ \\
peculiar proper &
$\mu_v=v/D$ &$\mu_{\alpha*,v}=v_\alpha/D$ & 
$\mu_{l*,v} = v_l/D$ \\
\phantom{MMM}motion & &$\mu_{\delta,v}=v_\delta/D$  & $\mu_{b,v}=v_b/D$ \\
& \multicolumn{3}{c}{$v_\perp=\sqrt{ {v_\alpha}^2+{v_\delta}^2}
= \sqrt{ {v_l}^2+{v_b}^2}$} \\
proper motion: & \multicolumn{3}{c}{$\mu_{\alpha*}=\mu_{\alpha*,G}+\mu_{\alpha*,v};\quad
\mu_\delta=\mu_{\delta,G}+\mu_{\delta,v}$}\\
\hline\hline
\multicolumn{4}{c}{\bf measured (nominal) values} \\
\hline
parallax, distance & $\varpi'$ & $D'=1/\varpi'$ \\
\hline
velocity$^a$ & $v'$ &  $v'_\alpha,v'_\delta,v'_r$ &
  $v'_l,v'_b,v'_r$ \\
proper motion$^a$ &
$\mu'$ &$\mu'_{\alpha*}=v'_\alpha/D$ & 
$\mu'_{l*} = v'_l/D$ \\
& &$\mu'_\delta=v'_\delta/D$  & $\mu'_b=v'_b/D$ \\
& \multicolumn{3}{c}{$v'_\perp=\sqrt{ {v'_\alpha}^2+{v'_\delta}^2}
= \sqrt{ {v'_l}^2+{v'_b}^2}$} \\
\hline
\end{tabular}
}
\tablefoot{$^a$ note that the measured values differ from the actual
values not only due to measurement error, but also due to correction
for galactic rotation}
\end{table}

\section{Data\label{s:data}}

To obtain a master list of pulsars with measured proper motions, 
we start by collating articles with proper motion measurements.
The ATNF Catalogue, version
1.54\footnote{www.atnf.csiro.au/research/pulsar/psrcat} 
(Manchester et al.\ 2005)\nocite{atnf}, was very helpful in this. 

Brisken et al.\ (2000)\nocite{brisken2000} note that VLBI measurements of proper motions
need to be corrected for ionospheric refraction. We therefore do not
use articles with proper motions from VLBI published before 2000.  To
select first-born, single pulsars we reject recycled
pulsars (i.e.\ those with $\dot P<5\times10^{-18}\mathrm{s\,s}^{-1}$),
pulsars in binaries, and pulsars in globular clusters.

In this first application of our new method we prefer to use
relatively accurate measurements.  We therefore omit pulsars with
distances determined only from dispersion measures, and pulsars with
proper motions determined from pulse timing. In both cases, the errors
are at least an order of magnitude larger than the errors obtained
with VLBI, and often only correspond to (upper or lower) limits. 
Distances from dispersion measures have uncertainties dominated
by systematic effects, with highly non-gaussian distributions. (For
pulsars distances and dispersion measures, see e.g.\ Yao et al.\ 2017;
\nocite{ymw2017} for proper motions from timing, see Hobbs et al.\
2004.) \nocite{hlk04}
We also omit proper motions of pulsars derived from displacements in
X-ray or optical images, which are relative to other
objects in the field of view. The conversion to absolute proper
motions in the ICRS adds significantly to the error.

\begin{table}
\caption{Sources for proper motions in our master list \label{t:sources}}
\centerline{
\begin{tabular}{c|llrr}
$S$ & \multicolumn{2}{c}{source}  & N & n \\
1 & Brisken et al.\ (2002) \nocite{bbgt02} & Table\,4 & 6  & 2 \\
2 & Brisken et al.\ (2003b) \nocite{btgg03}  & Table\,3 & 1  & 1 \\
3 & Chatterjee et al.\ (2001) \nocite{ccl+01} & Table\,2 & 1 & 1 \\
4 & Chatterjee et al.\ (2004) \nocite{ccv+04} & Table\,1 & 1 & 1\\
5 & Deller et al.\ (2009) \nocite{dtbr09} & Table\,3 & 4 & 2  \\
6 & Chatterjee et al.\ (2009) \nocite{cbv+09} & Table\,2 & 12 & 9 \\
7 & Kirsten et al. (2015) \nocite{kirsten_2015} & Table\,5 & 3 & 3 \\
     & & \hfill total: & 28 & 19 \\
 \hline
\end{tabular}
}
\tablefoot{$S$ source indicator, $N$ number of entries used (isolated pulsars with parallax measurements),
$n$ with age less than 10\,Myr. Later measurements may replace
earlier  ones; the source actually used is indicated in Table\,\ref{t:master}}
\end{table}

\begin{figure*}
\centerline{\includegraphics[width=14.cm]{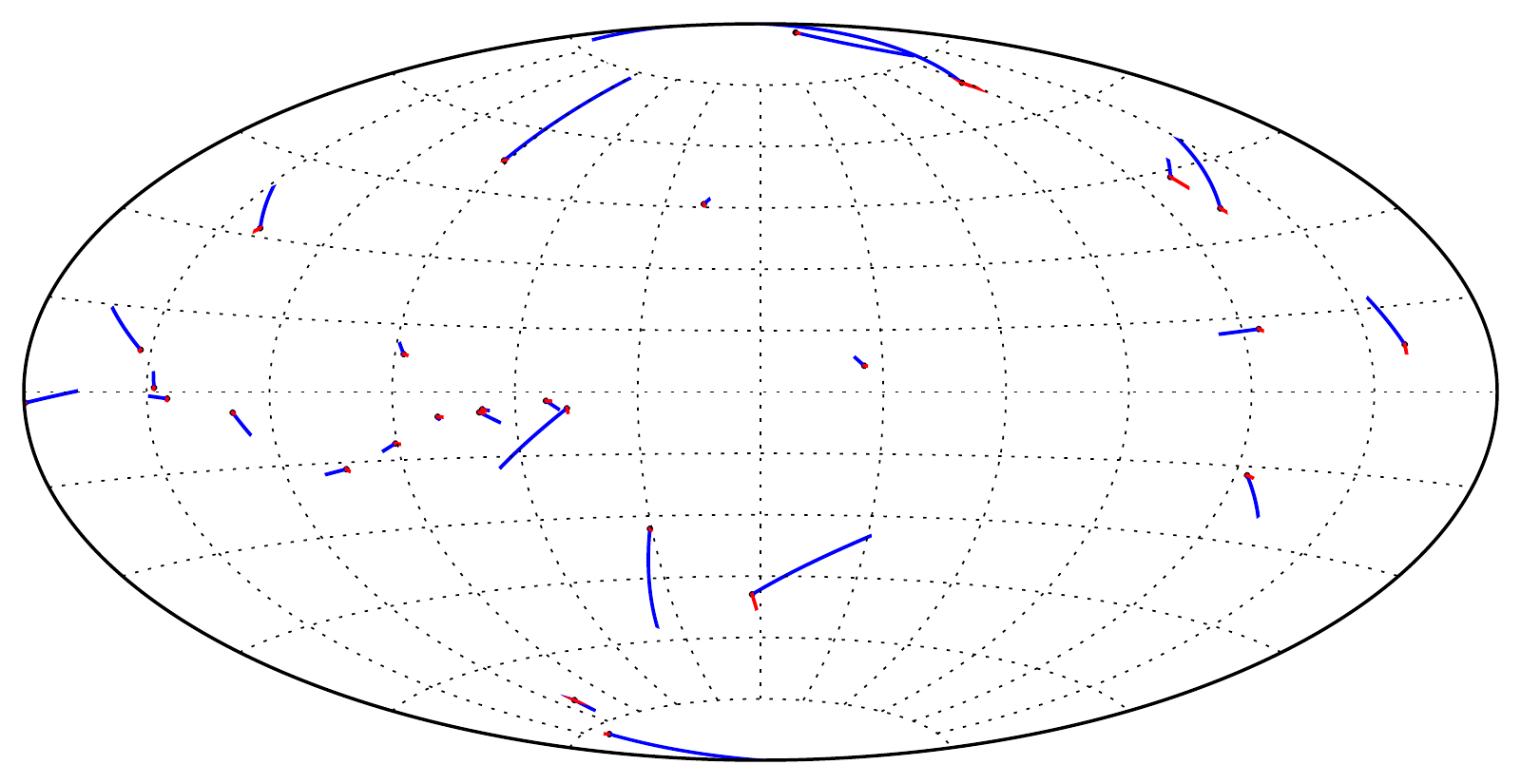}}
\centerline{\includegraphics[width=14.cm]{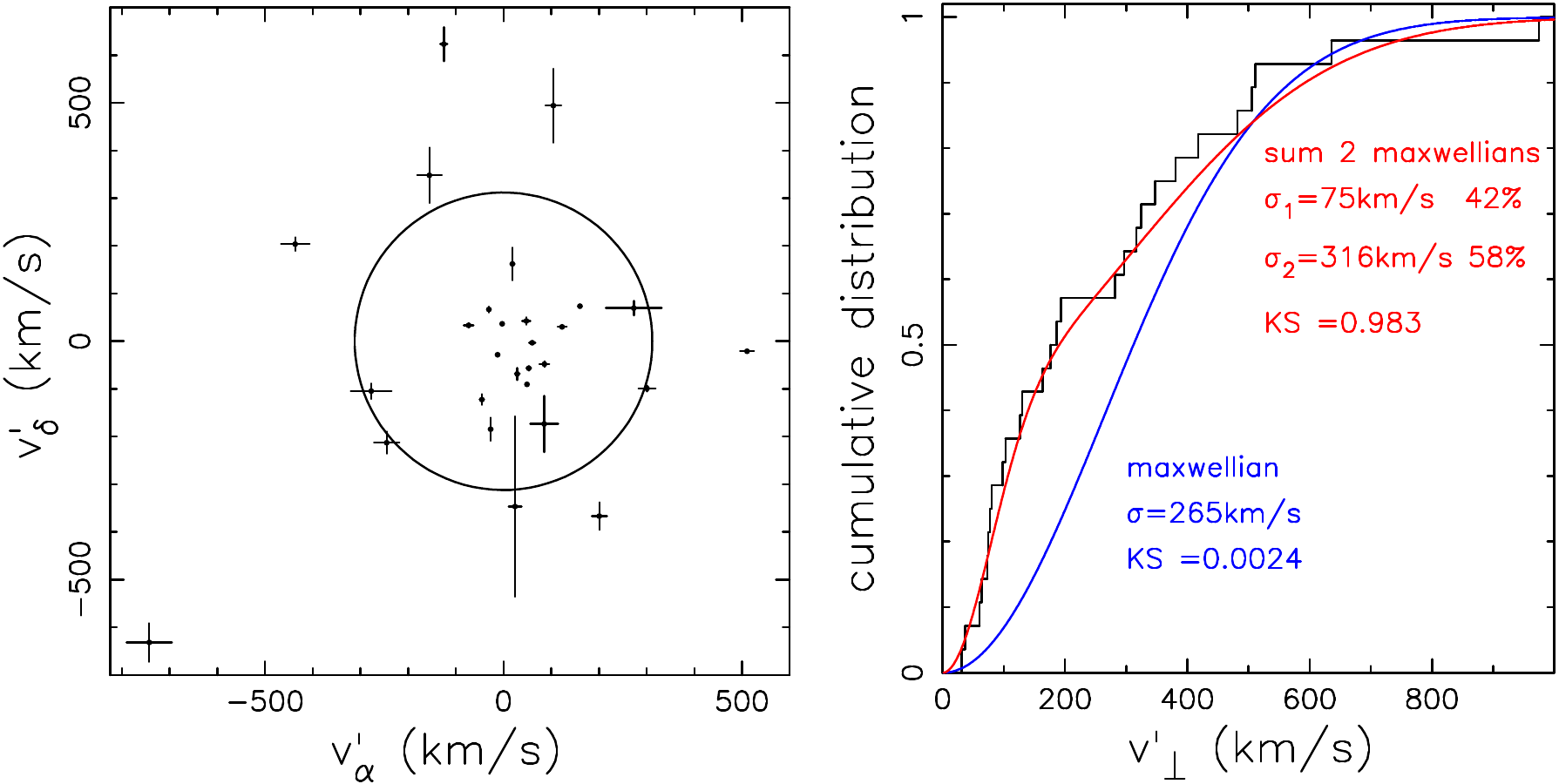}}
  \caption{Illustrations of data from our master list  of pulsars Table\,\ref{t:master}.
Top: celestial distribution in galactic coordinates.
The blue lines show the observed proper motion $\mu'$ and the red lines 
the correction due to galactic rotation (for nominal distance $D'$),
in 0.5 Myr. Below left: nominal velocities in the celestial plane. 
The circle indicates the median value for $v_\perp$ for the projection
of a maxwellian: $\sqrt{2\ln2}\,\sigma$, for $\sigma=265$\,km/s. 
Below right: cumulative distributions of the observed $v'_\perp$, and of 
$v_\perp$, blue:  according to Hobbs et al.\
(2005), red: according to our best solution, with the
$p$-value according to a one-sided Kolmogorov-Smirnov test that
the observed distribution is drawn from the theoretical one.
\label{f:observed}}
\end{figure*}

This leaves us with the VLBI measurements of the articles listed in
Table\,\ref{t:sources}.  Although the measurement of the proper motion
components are not independent of each other, the covariance value is
only provided by Brisken et al.\ (2003a) \nocite{bfg+03} who give no parallax values.  We
therefore ignore covariances between $\mu'_{\alpha*}$ and
$\mu'_\delta$.  In the majority of the measurements, the errors are
symmetric, and where asymmetric, the difference is small.  We simplify
our analysis by taking the largest error when errors are asymmetric.
(Test calculations in which the smallest error is taken give the same
results.)

The resulting master list of observed proper
motions in equatorial coordinates is given in Table\,\ref{t:master}.
Note that the proper motions in this Table are the observed
proper motions $\mu'_{\alpha*}$ and  $\mu'_{\delta}$,
not corrected for galactic rotation and peculiar solar velocity.
The celestial distribution, measured proper motions and nominal
velocities of the pulsars in our master list are illustrated in Fig.\ref{f:observed}.
From the top figure we learn that the correction for galactic motion in
general is small. The lower figures add to our suspicion that a
single maxwellian with $\sigma=265$\,km/s seriously underestimates the
number of pulsars with low velocities.


\section{Ingredients\label{s:ingredients}}

To determine the pulsar velocity distribution we use the measured
values of the parallax $\varpi'$ and of the two components of the proper motion
$\mu'_{\alpha*}$ and $\mu'_\delta$. 
The conditional probabilities of obtaining these measured values
when the actual values are $\varpi=1/D$, $\mu_{\alpha*}$ and $\mu_\delta$ 
can be written separately as
\begin{equation}
g_D(\varpi'|D)\Delta\varpi' = {1\over\sqrt{2\pi}\sigma_\varpi}
\exp\left[-\,{(1/D-\varpi')^2\over2{\sigma_\varpi}^2}\right] \Delta\varpi'
\label{e:erpar}\end{equation}
\begin{equation}
g_\alpha(\mu'_{\alpha*}|\mu_{\alpha*})\Delta\mu'_{\alpha*} = {1\over\sqrt{2\pi}\sigma_\alpha}
\exp\left[-\,{(\mu_{\alpha*}-\mu'_{\alpha*})^2\over2{\sigma_\alpha}^2}\right] \Delta\mu'_{\alpha*}
\label{e:ermua}
\end{equation}
\begin{equation}
g_\delta(\mu'_{\delta}|\mu_{\delta})\Delta\mu'_{\delta} = {1\over\sqrt{2\pi}\sigma_\delta}
\exp\left[-\,{(\mu_{\delta}-\mu'_{\delta})^2\over2{\sigma_\delta}^2}\right] 
\Delta\mu'_{\delta}
\label{e:ermud}
\end{equation}
where $\sigma_\varpi$, $\sigma_\alpha$ and $\sigma_\delta$ are the
measurement errors for the parallax and for the two components of the proper
motion, respectively. 

\begin{table}
\caption{Values of constants defining coordinate transformations and
velocity corrections  \label{t:constants}}
\centerline{
\begin{tabular}{cccl}
\multicolumn{3}{c}{Galactic pole, longitude node}  &  \\
$\alpha_\mathrm{GP} = 192\fdg85948$ &
$\delta_\mathrm{GP} = 27\fdg12825$ & $l_\Omega = 32\fdg93192$ &
a \\
\hline
\multicolumn{3}{c}{Peculiar velocity Sun}  &  \\
$U=10.0$\,km/s & $V=5.3$\,km/s & $W=7.2$\,km/s & b \\
\hline
\multicolumn{3}{c}{Galactic rotation}  &  \\
\multicolumn{2}{c}{$v_R(R_o) = v_R(R) = 220\,\mathrm{km/s}$}  & & b
  \\
\hline
\multicolumn{3}{c}{Distance galactic center, scales pulsar distribution} \\
$R_o=8.5$\,kpc & $h = 0.33$\, kpc & $H = 1.7$\, kpc  & c\\
\end{tabular}
}
\tablefoot{For explanation of these constants see
Appendices\,\ref{s:conversions} and \,\ref{s:correct}.\\
a: from Perryman et al.\ 1997\nocite{perryman1997},  
b: from Dehnen \& Binney 1998\nocite{dehnen1998}, 
c: from Verbiest et al. (2012).
}
\end{table}

To obtain the joint probability of the measured and actual values,
these equations must be complemented with the equations indicating
the probability density functions of the actual distance and proper
motion. 

The probability density $f_D(D)$  of the distance $D$ of the pulsar to the Earth
for a galactocentric pulsar distribution is 
given by Verbiest et al.\ (2012).\nocite{vwc+12} In the notation of Igoshev et al.\ (2016):
\begin{equation}
f_D(D)\propto D^2R^{1.9}\exp\left[-{|z(D,b)|\over h}-{R(D,l,b)\over H}\right]
\equiv D^2\mathcal{F}(D)
\label{e:fd}
\end{equation}
with
\begin{equation}
z = D\sin b;\,\mathrm{and}\, R=\sqrt{{R_o}^2+(D\cos b)^2-2D\cos b\,R_o\cos l}
\end{equation}
where $R$ and $R_o$ are the galactocentric distance of the pulsar and
the Sun, respectively, projected on the galactic plane.
Through $\mathcal{F}(D)$ also $f_D(D)$ is a function of  galactic coordinates $l,b$.

The proper motion of a pulsar $\mu_{\alpha*},\mu_\delta$ is the sum of the
proper motion of its standard of rest with respect to the Sun
$\mu_{\alpha*,G},\mu_{\delta,G}$  and the proper motion
caused by its velocity with respect to its local
standard of rest $\mu_{\alpha*,v},\mu_{\delta,v}$:
\begin{equation}
\mu_{\alpha*}=\mu_{\alpha*,G}+\mu_{\alpha*,v};\quad
\mu_\delta=\mu_{\delta,G}+\mu_{\delta,v}
\label{e:mucor}\end{equation}
The derivation of $\mu_{\alpha*,G}$ and $\mu_{\delta,G}$ is described in Appendices
\ref{s:conversions} and \ref{s:correct}. 
The velocity of the local standard of rest is assumed to be the galactic rotation
velocity, $v_R(R_o)$ for the Sun and $v_R(R)$ for the pulsar.
The peculiar velocity of the Sun is [$U$,$V$,$W$],
where the components are respectively in the direction from the Sun
towards the galactic centre, in the direction of the galactic
rotation, and perpendicular to the galactic plane.
In galactic coordinates 
\begin{equation}
D\,\mu_{l*,G}=U\sin l-[V+v_R(R_o)]\cos l +v_R(R)\cos(\theta+l)
\label{e:vsubl}\end{equation}
and 
$$ D\,\mu_{b,G} = \Big[U\cos l+[V+v_R(R_o)]\sin l
-v_R(R)\sin(\theta+l)\Big]\sin b\phantom{and} $$
\begin{equation}
\phantom{oliebollen} -W\cos b 
\label{e:vsub}\end{equation}
The angle $(\theta+l)$ is computed from:
\begin{equation}
\tan(\theta+l) = {R_o\sin l\over R_o\cos l-D\cos b} 
\label{e:theta}\end{equation}
The values for [$U$,$V$,$W$], $v_R$ and $R_o$ that we use are listed
in Table\,\ref{t:constants}. To compare velocities expressed in km/s
with proper motions expressed in mas/yr, we use the conversion
\begin{equation}
v\mathrm{(km/s)} = 4.74\,\mu\mathrm{(mas/yr)}\,D\mathrm{(kpc)}
\end{equation}

The pair $\mu_{l*,G},\mu_{b,G}$, is converted to the pair in equatorial coordinates
$\mu_{\alpha*,G},\mu_{\delta,G}$
with the rotation given by Eqs.\ref{e:rotcora},\ref{e:rotcorb}.
Note that $\mu_{l*,G}$ and $\mu_{b,G}$ depend on the (unknown) distance.
This is the reason that Table\,\ref{t:master} gives the observed
proper motions, not corrected for galactic rotation and solar motion.

$\mu_{\alpha*,v}$ and $\mu_{\delta,v}$ depend on the peculiar velocity
$v$ of the pulsar and on the direction of this velocity. 

\begin{figure*}
\centerline{\includegraphics[width=\columnwidth]{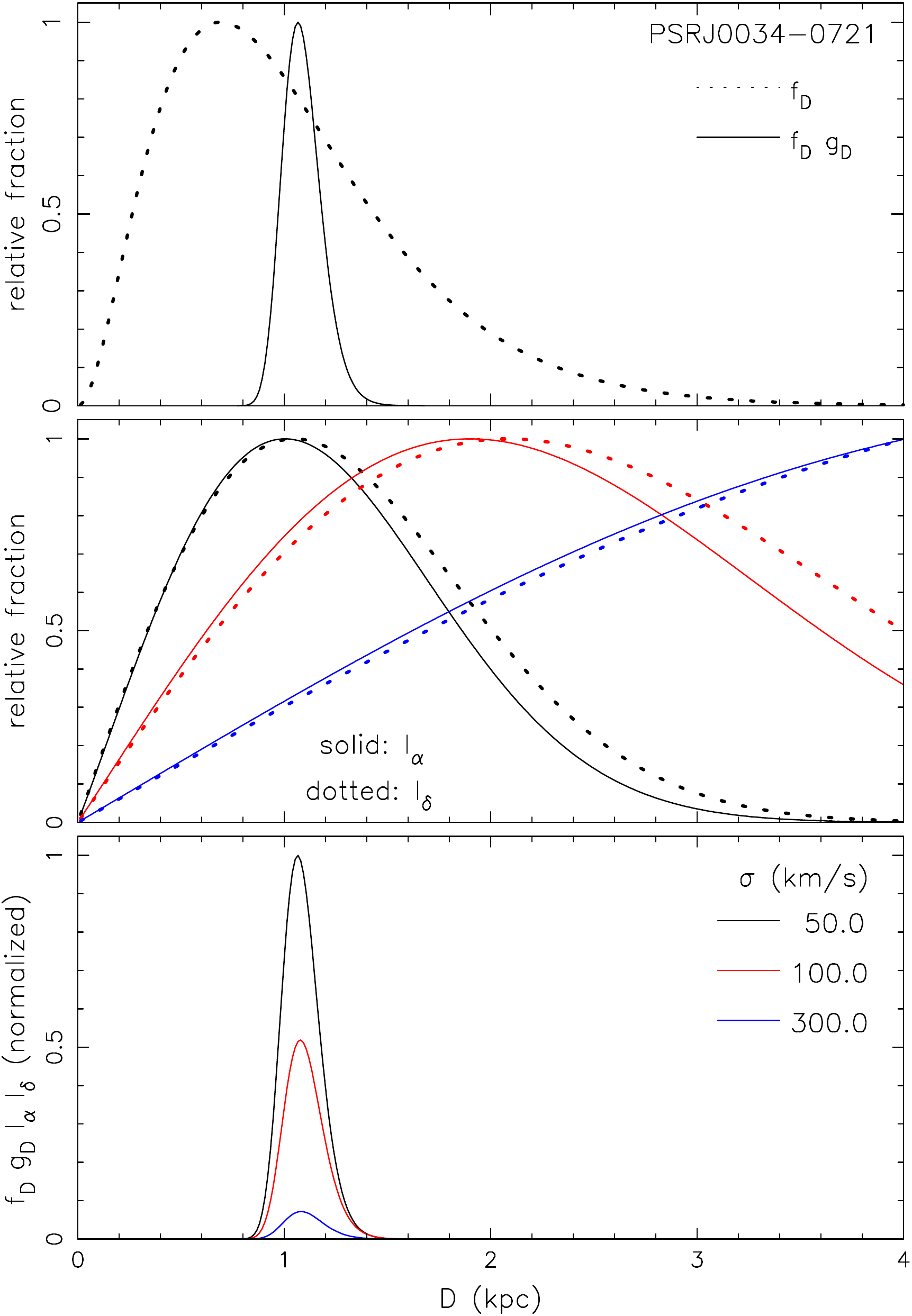}\hfill
\includegraphics[width=\columnwidth]{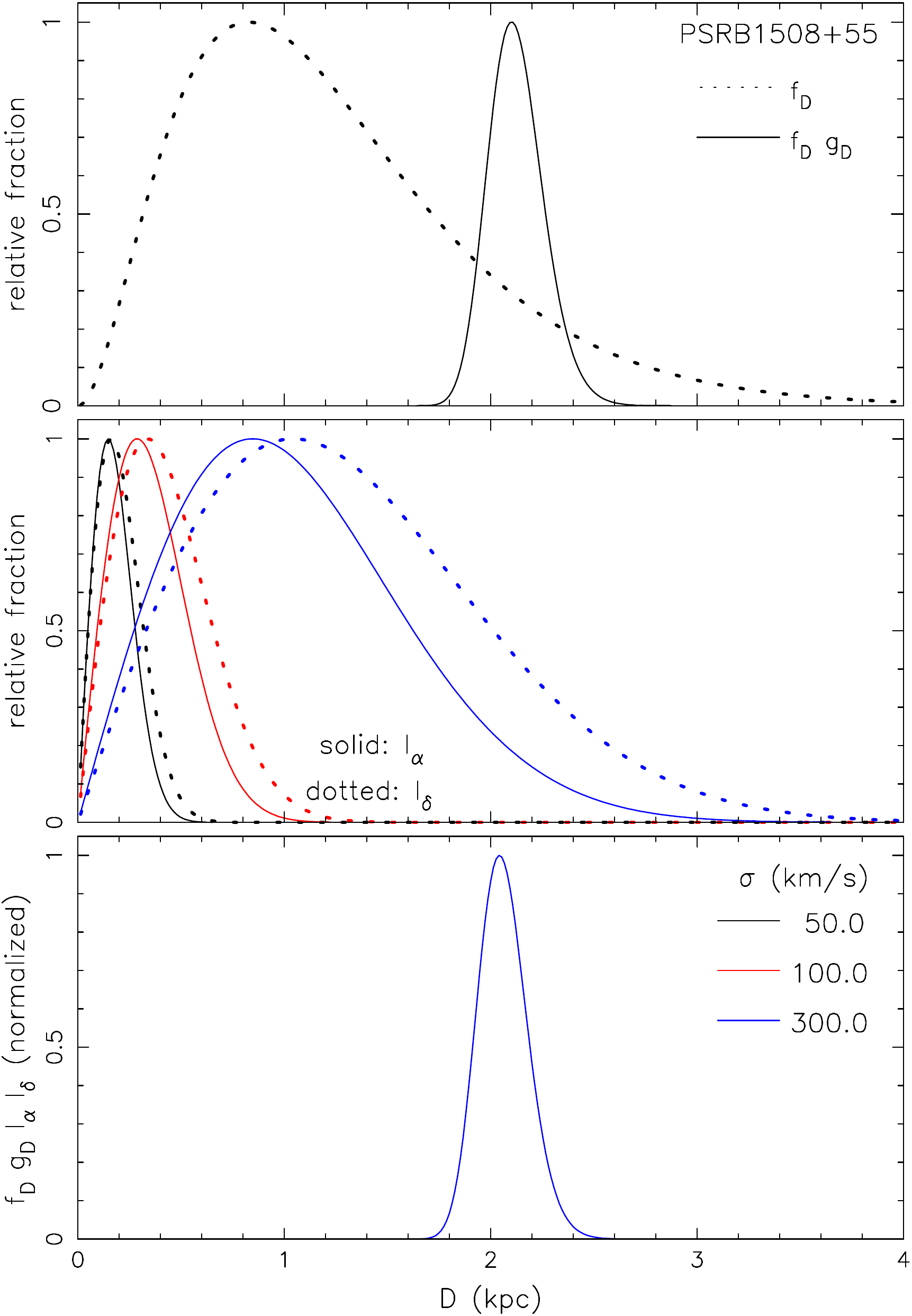}}
\caption{Illustration for two pulsars of the contributions  to the integrand of the likelihood 
 $L_\mathrm{maxw}(\sigma)$ (Eq.\,\ref{e:fin_maxw}) of the separate factors
  $f_D(D)g_D(\varpi'|D)$ (top graphs; Eqs.\,\ref{e:erpar}, \ref{e:fd}),
  $I_\alpha$ and $I_\delta$ (middle graphs). Each curve has been normalized separately 
to maximum unity. $I_\alpha$ and $I_\delta$ are shown for for three 
values $\sigma$ of the maxwellian (Eq.\,\ref{e:maxw}). The lower 
graphs show the integrand of the likelihood $L_\mathrm{maxw}(\sigma)$
for three values of $\sigma$, as a function of distance, normalized
to the highest maximum of the three. The measurements of
PSR\,J0034$-$0721 (left) favour a low value of $\sigma$. Those of 
\psrbad\ (right) require a high value of $\sigma$, and the integrands for
$\sigma=50$ and $100$\,km/s are indistinguishable from zero in this graph.
\label{f:examples}}
\vspace*{0.4cm}
\end{figure*}

\subsection{Best solution and fiducial intervals}

In the following sections we will discuss a number of models, and
for each model compute a likelihood $L_i(\vec\sigma)$ for an
individual pulsar labelled $i$, as a function of the parameter vector
$\vec\sigma$. We then construct the deviance $\mathcal{L}$ with
\begin{equation}
\mathcal{L}(\vec\sigma) = 
-2\sum_{i=1}^N \ln L_i(\vec\sigma)
\label{e:deviance}\end{equation}
where $N$ is the number of pulsars.
$\vec\sigma_\mathrm{opt}$ is the parameter vector for which
Eq.\,\ref{e:deviance} reaches its minimum.
We write differences with the optimal solution as
\begin{equation}
\Delta\mathcal{L}(\vec\sigma) \equiv
\mathcal{L}(\vec\sigma) - \mathcal{L}(\vec\sigma_\mathrm{opt}) 
\label{e:dfitness}\end{equation}
For appropriate choices of $L_i$ these differences
approximate a $\chi^2$ distribution. 
For a parameter vector consisting of a single parameter,
we estimate its 68\%\ range by determining for which
values Eq.\,\ref{e:dfitness} is equal to 1.
To determine the range of values
if the vector parameter has three parameters, we proceed as follows.
We fix the value of one parameter at an offset from the optimal
value, and then determine the combination of the two other
parameters that gives the lowest value for
$\Delta\mathcal{L}(\vec\sigma)$.
We vary the offset until this lowest value is 1.
Repeating this for each of the three parameters for positive and
negative offsets from the best values gives the ranges listed in
Table\,\ref{t:results}.

Note that the best parameter values and the fiducial ranges determined 
this way do not depend on the normalization of $L_i$: a constant 
multiplicative factor $x$ to any $L_i$ leads to a constant additive factor
$-2\ln x$ in Eq.\,\ref{e:deviance} and drops out in Eq.\,\ref{e:dfitness}.

We will also use the deviance to compare different models, using
\begin{equation}
d\mathcal{L}\equiv
\mathcal{L}^a({\vec\sigma_\mathrm{opt}}^a) - \mathcal{L}^b({\vec\sigma_\mathrm{opt}}^b) 
\label{e:diffitness}\end{equation}
where indices $a$ and $b$ refer to the different
models. The distribution of $d\mathcal{L}$ approximates a $\chi^2$
distribution less well than $\Delta\mathcal{L}$, but we will use this
difference as a rough indication of relative merit of models.

\section{Maxwellian velocity distribution\label{s:maxwell}}

\begin{figure}
\centerline{\includegraphics[width=\columnwidth]{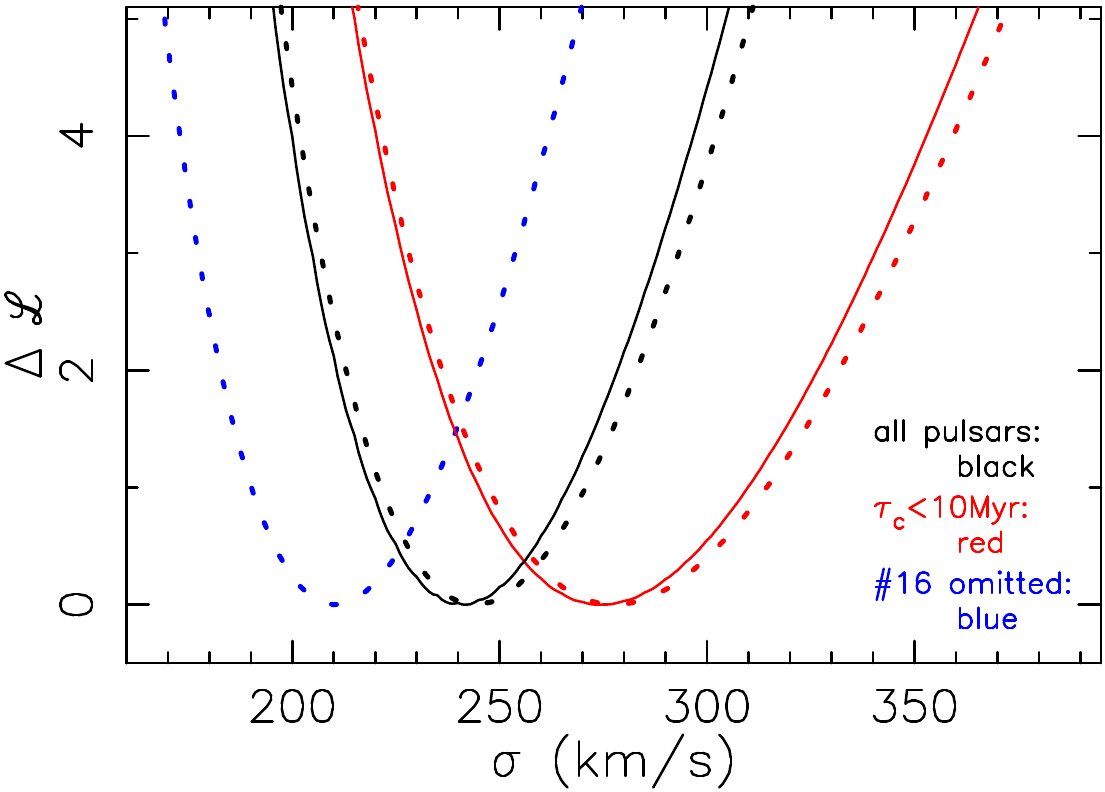}}
\caption{Variation of $\mathcal{L}$ with velocity distribution
  parameter $\sigma$ for the model with a single isotropic maxwellian
  (dotted lines), and for the mixed model in which most pulsars
 have an assumed semi-isotropic velocity distribution (solid lines).
 The colour coding indicates the pulsar sample:
 all 28 pulsars in our master list,  the 27 pulsars remaining
 after removing \psrbad, and the 19 youngest ($\tau<10$\,Myr) pulsars.
 \label{f:singlemax}}
\end{figure}

The maxwellian velocity distribution is characterised by
a single parameter $\sigma$:
\begin{equation}
f(v,\sigma)dv = \sqrt{2\over\pi}
{v^2\over\sigma^3}\exp\left[-{v^2\over2\sigma^2}\right]dv; 
\quad (0<v<\infty)
\label{e:maxw}
\end{equation}
In the isotropic case, the maxwellian may
be decomposed in three independent gaussians in any three
mutually perpendicular directions. We choose the directions
of increasing right ascension, increasing declination, and
the radial direction. This enables us to write
the joint probability of measured values $\varpi'$, $\mu'_{\alpha*}$,
$\mu'_\delta$ and actual values $D$, $v_\alpha=D\mu_{\alpha*,v}$ and
$v_\delta=D\mu_{\delta,v}$ as
$$
P_\mathrm{maxw}(\varpi',\mu'_{\alpha*},\mu'_\delta,D,v_\alpha,v_\delta,v_r)=
G(v_\alpha,\sigma)G(v_\delta,\sigma)G(v_r,\sigma)$$
\vspace*{-0.5cm}
\begin{eqnarray}
\phantom{oliebolb}&\times&{f_D(D)\over\int_0^{D_\mathrm{max}}f_D(D)dD}\,
{1\over\sigma_\varpi\sqrt{2\pi}}\exp\left[-\,{(1/D-\varpi')^2\over2{\sigma_\varpi}^2}\right] \nonumber \\
&\times&{1\over\sigma_\alpha\sqrt{2\pi}}\exp\left[-\,{(\mu_{\alpha*,G}(D)+v_\alpha/D
-\mu'_{\alpha*})^2\over2{\sigma_\alpha}^2}\right] 
\nonumber \\
&\times&{1\over\sigma_\delta\sqrt{2\pi}}\exp\left[-\,{(\mu_{\delta,G}(D)+v_\delta/D
-\mu'_{\delta})^2\over2{\sigma_\delta}^2}\right] 
\label{e:definejm}\end{eqnarray}
where
\begin{equation}
G(v,\sigma) =
{1\over\sigma\sqrt{2\pi}}\exp\left[-{v^2\over2\sigma^2}\right];
\quad (-\infty<v<\infty)
\end{equation}
To obtain the value of $\sigma$ which gives the most likely
correspondence with the measurements, we must take into account
contributions to the likelihood of all distances and velocities.
We therefore define the likelihood for the maxwellian as
\begin{equation}
L_\mathrm{maxw}(\sigma) =
\int_o^{D_\mathrm{max}}\int_{-\infty}^\infty\int_{-\infty}^\infty\int_{-\infty}^\infty
 P_\mathrm{maxw}dv_\alpha dv_\delta dv_r dD
\label{e:jointmaxwel}\end{equation}
The radial velocities occur only in $G(v_r,\sigma)$, and thus the integral
over $v_r$ can be computed separately: $\int_{-\infty}^\infty G(v_r,\sigma)dv_r=1$.
The integrals over $v_\alpha$ and $v_\delta$ are more involved, but can also
be solved analytically. Thus, for $v_\alpha$
$$\int_{-\infty}^\infty \exp\left\{-{1\over2}\left[{{v_\alpha}^2\over\sigma^2}
 +{\Big(v_\alpha+D(\mu_{\alpha*,G}-\mu'_{\alpha*})\Big)^2\over D^2{\sigma_\alpha}^2}
\right] \right\} dv_\alpha = \phantom{oliebolbakker}$$
\begin{equation}
\phantom{oliebol}
\sqrt{2\pi}\,\left({1\over\sigma^2}+{1\over D^2{\sigma_\alpha}^2}\right)^{-1/2}
\exp\left[-{1\over2}{D^2(\mu_{\alpha*,G}-\mu'_{\alpha*})^2\over
  \sigma^2+D^2{\sigma_\alpha}^2}\right]
\end{equation}
and analogously for $v_\delta$.
Taken together these results lead to
\begin{equation}
L_\mathrm{maxw}(\sigma)=\mathcal{C}\int_0^{D_\mathrm{max}}
f_D(D)\exp\left[-\,{(1/D-\varpi')^2\over2{\sigma_\varpi}^2}\right]\mathcal{I}_\alpha\mathcal{I}_\delta
dD
\label{e:fin_maxw}
\end{equation}
where
\begin{eqnarray}
\phantom{mmm}
\mathcal{C} &\equiv &
\left[(2\pi)^{3/2}\sigma_\varpi\sigma_\alpha\sigma_\delta
\int_0^{D_\mathrm{max}} f_D(D) dD\right]^{-1}\nonumber\\
\mathcal{I}_\alpha & \equiv &\left(1+{\sigma^2\over D^2{\sigma_\alpha}^2}\right)^{-1/2}
\exp\left[-{1\over2}{(D\,\mu_{\alpha*,G}-D\,\mu'_{\alpha*})^2\over
  \sigma^2+D^2{\sigma_\alpha}^2}\right] \nonumber \\
\phantom{mmm}
\mathcal{I}_\delta& \equiv &
\left(1+{\sigma^2\over D^2{\sigma_\delta}^2}\right)^{-1/2}
\exp\left[-{1\over2}{(D\,\mu_{\delta,G}-D\,\mu'_{\delta})^2\over
  \sigma^2+D^2{\sigma_\delta}^2}\right]\nonumber
\end{eqnarray}

The integral over distances in Eq.\,\ref{e:fin_maxw} is computed
numerically, out to $D_\mathrm{max}=10$\,kpc. 
$\mathcal{C}$ ensures that each distribution in Eq.\,\ref{e:definejm} 
is normalized to unity; in the computations $\mathcal{C}$
may be ignored, as it only adds a constant in the
deviance (Eq.\,\ref{e:deviance}) and drops out in Eq.\,\ref{e:dfitness}.

To illustrate the effect of the various factors in the integrand of
Eq.\,\ref{e:fin_maxw} we show these separately in
Fig.\,\ref{f:examples}, for two pulsars.  For a fixed velocity, the
proper motion scales inversely with the distance.  The large parallax
of PSR\,0034$-$0721, combined with its relatively small proper motion,
favours a maxwellian with a small average velocity, but still allows a
maxwellian with a high average velocity as this has a finite tail at
low velocities. In contrast, the smaller parallax of \psrbad\ combined
with its large proper motion, demands a maxwellian with a large
average velocity, because high velocities have vanishingly low
probability in a maxwellian with a low average velocity.

Labelling the likelihoods of Eq.\,\ref{e:fin_maxw} for each of $N$
pulsars with $i$, we compute the deviance $\mathcal{L}$ with
Eq.\,\ref{e:deviance}.
$\Delta\mathcal{L}(\sigma)$ (Eq.\,\ref{e:dfitness}) is shown for three
pulsar samples in Figure\,\ref{f:singlemax}.
The sample of all 28 pulsars in our master list (Table\,\ref{t:master})
leads to $\sigma_\mathrm{opt}\simeq 244$\,km/s, with a range of
about 50\,km/s found from $\Delta\mathcal{L}=1$;
see Table\,\ref{t:results}.
To illustrate the influence of a single pulsar, we also show
$\Delta\mathcal{L}(\sigma)$ for the sample of 27 pulsars remaining
after removing \psrbad, the pulsar with the worst likelihood 
for $\sigma=245$\,km/s. This sample has $\sigma_\mathrm{opt}\simeq210$\,km/s.
The reason for this shift is evident from Fig.\,\ref{f:examples}: the
measurements of  \psrbad\ require a large value of $\sigma$.
Removing any one of the 27 other pulsars from the full sample leads to a much
smaller shift.

The pulsar velocities of young pulsars, less affected by acceleration
in the galactic gravitational field, are more indicative of the pulsar
velocities at birth, and therefore we also investigate the sample of
the 19 youngest pulsars with characteristic age $\tau_c< 10$\,Myr.  This
leads to a higher optimal distribution parameter
$\sigma_\mathrm{opt}\simeq280$\,km/s. The smaller
number of pulsars also leads to a wider range
of $\sigma$ for which $\Delta\mathcal{L}(\sigma)<1$.
An upper limit to $\tau_c$ of 5\,Myr leads to the same
$\sigma_\mathrm{opt}$ as for 10\,Myr, but further widens the
uncertainty range.
Removing \psrbad\ from the sample of young pulsars
reduces the optimal distribution sample to 
 $\sigma_\mathrm{opt}\simeq235$\,km/s.

Figures\,\ref{f:observed} and \ref{f:examples} indicate that a single
maxwellian is not a good description of the velocity distribution
of young radio pulsars. 
We are therefore not unduly worried about the shifts in
$\sigma_\mathrm{opt}$ between the different samples, but move
on to investigate more promising models.

\section{Sum of two maxwellians\label{s:two_maxw}}

\begin{figure}
\centerline{\includegraphics[width=0.8\columnwidth]{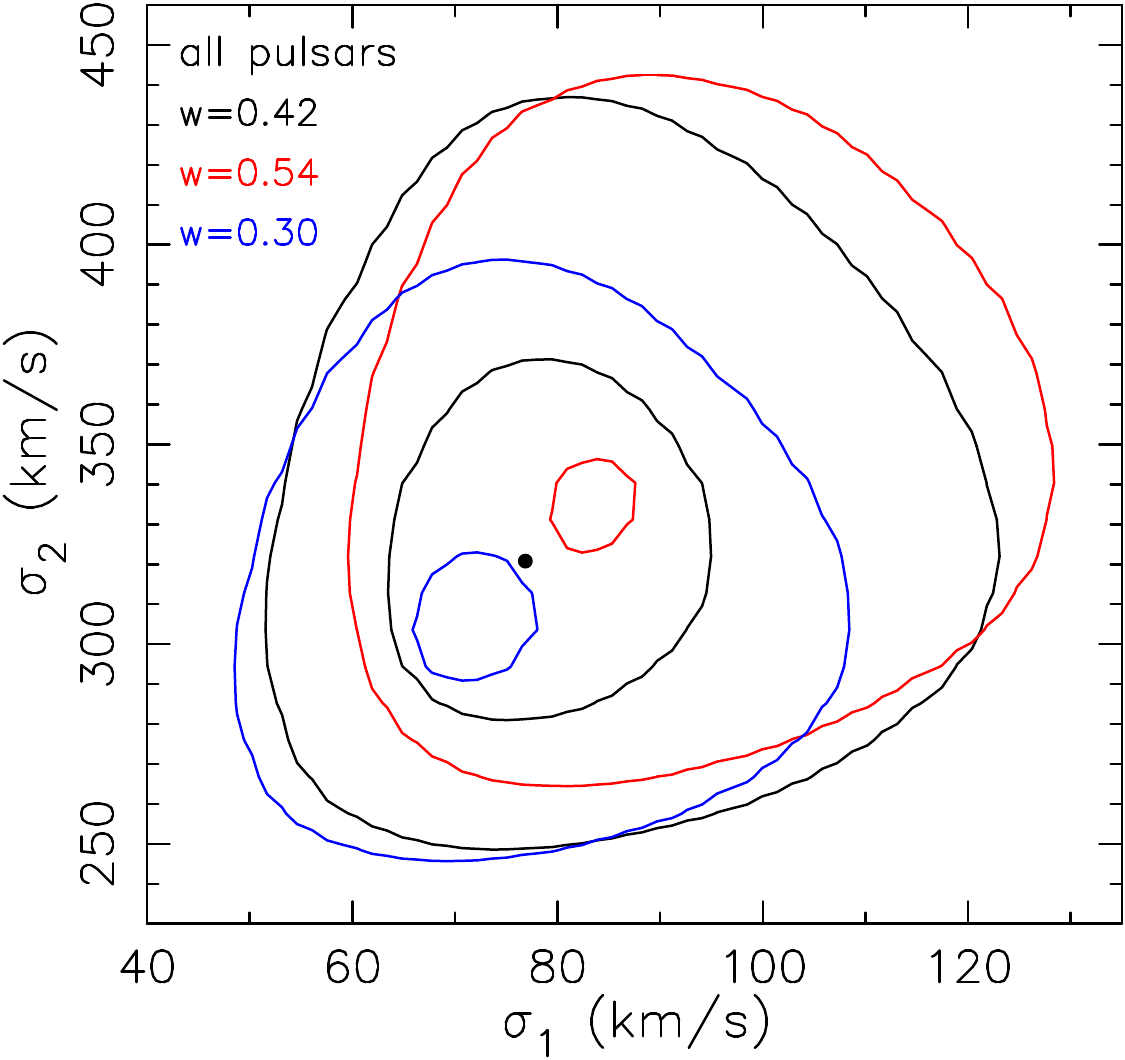}}
\centerline{\includegraphics[width=0.8\columnwidth]{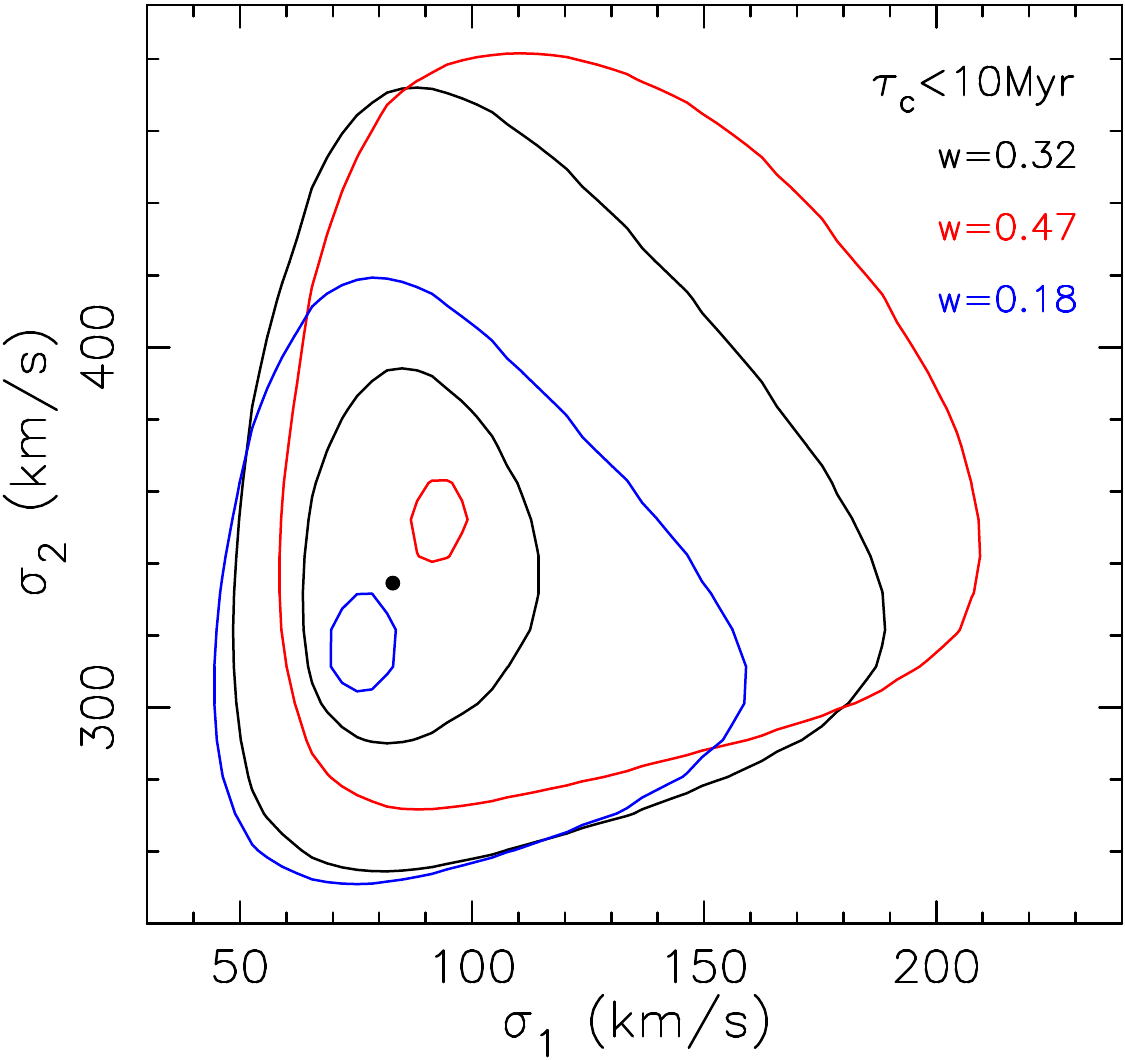}}
\caption{Contours of $\mathcal{L}(\vec\sigma)$ in three
    $\sigma_1,\sigma_2$ planes with fixed $w$, for the model
   with two isotropic maxwellians. 
   Contours of constant $\Delta\mathcal{L}(\vec\sigma)$
   (Eq.\,\ref{e:dfitness}) are shown for values 1 and 4, in each plane,
   The best solution is indicated with $\bullet$.
   Top: all pulsars. $\vec{\sigma}_\mathrm{opt} = $($77$km/s, $321$km/s, 0.42).
   Below: pulsars with $\tau_c<10$\,Myr.
   $\vec{\sigma}_\mathrm{opt} = $($83$km/s, $335$km/s, 0.32).
 \label{f:twomaxw}}
\end{figure}

We  investigate a velocity distribution which is
the sum of two maxwellians, one to explain the lower observed velocities, and
one for the higher velocities.
Defining the  vector of parameters $\vec\sigma=[\sigma_1, \sigma_2,w]$
we write
\begin{equation}
f_v (v, \vec\sigma) dv = \sqrt {\frac{2}{\pi}} v^2
\left[\frac{w}{\sigma_1^3} \exp \left( -\frac{1}{2}
    \frac{v^2}{\sigma_1^2} \right) 
 + \frac{(1-w)}{\sigma_2^3}  \exp \left( -\frac{1}{2} \frac{v^2}{\sigma_2^2} \right) \right] dv
\label{e:twomaxw}\end{equation}
The likelihood for the sum of two maxwellians is the sum of the likelihoods 
of the two maxwellians: in analogy with Eq.\,\ref{e:fin_maxw} we have
\begin{equation}
L_{2\mathrm{maxw}}(\vec\sigma) = w L_\mathrm{maxw}(\sigma_1) + (1-w) L_\mathrm{maxw}(\sigma_2)
\label{eq:2maxw}
\end{equation}
We compute $L_\mathrm{maxw}(\sigma)$ on a grid of values of $\sigma$,
in steps of 1\,km/s, and use the subroutine {\tt AMOEBA} of Press et
al.\ (1986) \nocite{press86}, which implements the downhill simplex method
of Nelder and Mead, to obtain the optimal values of $w$, $\sigma_1$ and
$\sigma_2$ for which $\mathcal{L}$, computed from Eq.\,\ref{eq:2maxw}
with Eq.\,\ref{e:deviance}, has its minimum.
The results are listed in Table\,\ref{t:results}, and illustated in
Figs.\,\ref{f:twomaxw}.

To decide on the significance of the second maxwellian, we note
that it adds two parameters to the model with one maxwellian,
and compute the deviance difference $d\mathcal{L}$ with Eq.\,\ref{e:diffitness}.
We first investigate the sample of all 28 pulsars in our master list
(sample A).  For this sample, $d\mathcal{L} =-14$,
indicating that the addition of a second maxwellian is very
significant ($\Delta\chi^2=-14$ corresponds to a 99.8\%\ confidence
level for 2 added parameters).
The low-velocity component represents between 29\%\ and 54\%\ of 
the pulsar population. Fig.\,\ref{f:twomaxw} shows that 
the values of $\sigma_1$ and $\sigma_2$ are mildly correlated with
$w$: a larger (smaller) fraction of the low-velocity component
leads to larger (smaller) values of $\sigma_1$ and $\sigma_2$.
The shift, however, lies well within the error range of
$\sigma_1$ and $\sigma_2$; the main effect of the correlation
between $\sigma_1$ and $\sigma_2$ is to  mitigate the drop
of pulsar numbers with velocities between $\sigma_1$ and $\sigma_2$.

The sample of 19 pulsars in our list with characteristic age 
$\tau_c<10$\,Myr (sample Y) leads to the same result, but with
somewhat lager error margins for the parameters $\vec\sigma$.
For these young pulsars, the evidence for a second maxwellian is
still significant ($\Delta\chi^2=-6$ is 95\%\ confidence).

\begin{figure}
\centerline{\includegraphics[width=\columnwidth]{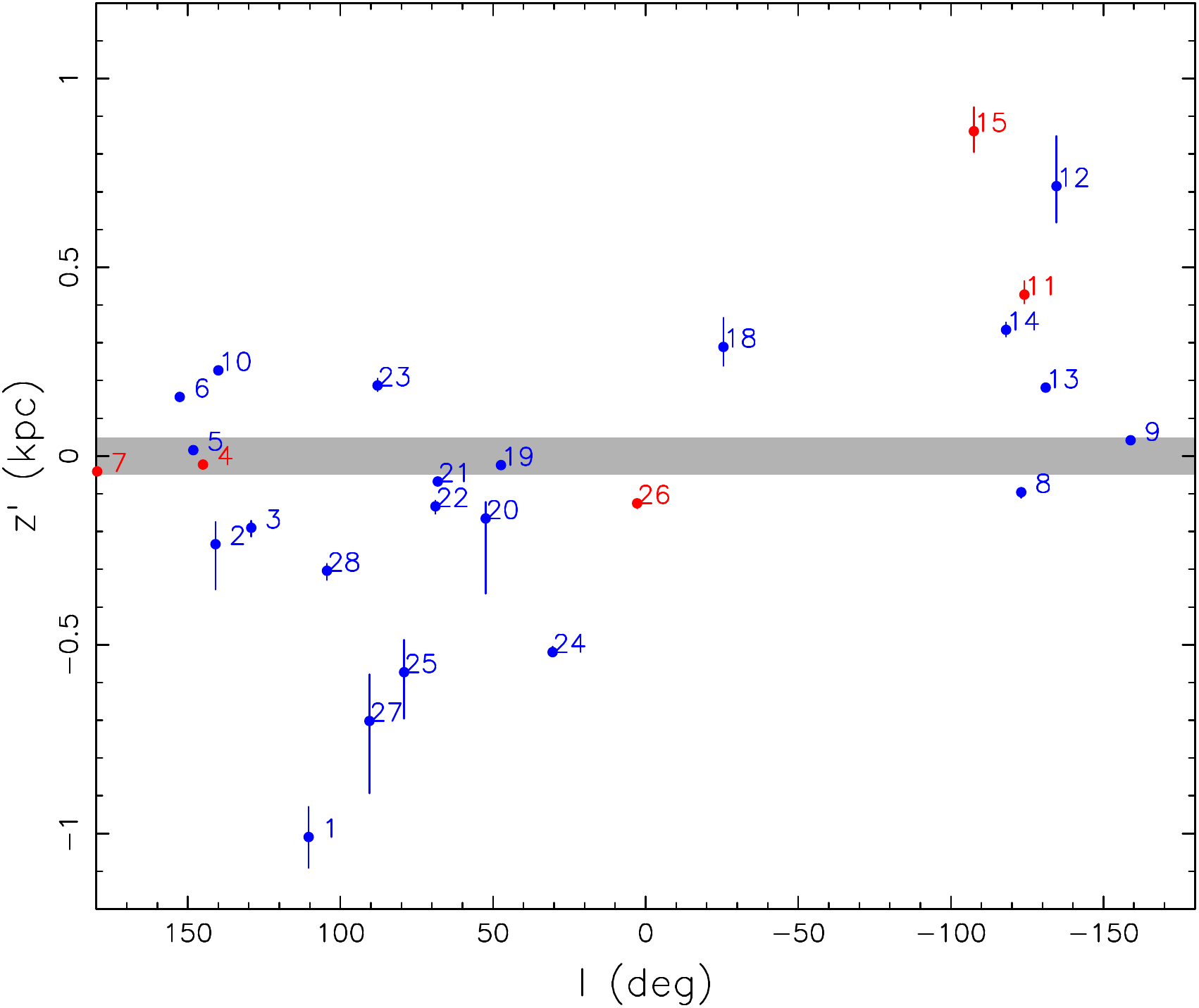}}
\caption{Nominal distance from the Galactic plane $z'=\sin b/\varpi'$,
range $\sin b/(\varpi'\pm\sigma_\varpi)$, as a 
function of longitude. The blue points indicate pulsars nominally 
moving away from the plane, i.e.\ $z'$ and $v'_z=\mu'_b\cos b/\varpi'$ have the
same sign; the red points are pulsars nominally moving towards the
plane. The grey band indicates the scale height of 50\,pc of O-stars.
The numbers refer to the sequence
number in Table\,\ref{t:master}. Numbers 16  at $z'=1.6$\,kpc and 17 at $5.5$\,kpc,
respectively, are outside the frame, and both are moving away from 
the galactic plane.}
\label{f:lzvz}
\end{figure}

\section{Semi-isotropic maxwellian velocity distribution\label{s:semiiso}}

\begin{table*}
\caption{Results of the model calculations for all 28
 pulsars in our master list (A), and for the 19 youngest pulsars
 ($\tau_c<10$\,Myr, Y). \label{t:results}}
\center{
\begin{tabular}{|l|c@{\hspace{0.2cm}}c|r@{\hspace{0.3cm}}c@{\hspace{0.2cm}}r|r@{\hspace{0.2cm}}cr@{\hspace{0.2cm}}cr@{\hspace{0.2cm}}cc||r@{\hspace{0.3cm}}c|}
\hline
& \multicolumn{2}{c|}{sample}   & \multicolumn{3}{c|}{single maxwellian}
& \multicolumn{7}{c||}{two maxwellians} & \multicolumn{2}{c|}{$v_l$ Gaussian}\\
\hline
&   & $N$ & $\sigma$ & range  & $d\mathcal{L}$  & $\sigma_1$ & range &
   $\sigma_2$ & range & $w$ & range & $d\mathcal{L}$ & $\sigma$ & range \\
 &  & & \multicolumn{2}{c}{(km/s)} & & \multicolumn{2}{c}{(km/s)} & \multicolumn{2}{c}{(km/s)} &
   \multicolumn{2}{c}{(\%)} & &  \multicolumn{2}{c|}{(km/s)} \\
\hline
isotropic models &A & 28 & 244  & 221-271 & $\equiv$0 &  77 & 62-97 &
321 & 278-375 & 42 & 29-54 & $-$14 & 240 & 209-279\\
mixed models & A & 22+6  & 239 & 219-267 & $-$18 & 75 & 61-95 & 316 &
276-369 & 42 & 30-55 & $-$33 & & \\
\hline
isotropic models &Y & 19 & 277 & 247-314 & $\equiv$0 & 83 &  62-117  &
335 & 287-398 & 32 & 17-47 & $-$6 & 263 & 223-314\\
mixed models & Y & 14+5 & 273 & 245-310 & $-$16 & 82 & 61-115 & 328 &
285-391 & 32 & 17-48 & $-$22 & & \\
\hline
\end{tabular}}
\tablefoot{For the mixed models we give separately the number
  of pulsars from a semi-isotropic and an isotropic distribution (see Table\,\ref{t:master}).
  For each model we give the best parameters and their approximate 68\%\ range
  determined by setting Eq.\,\ref{e:dfitness} 
  to unity.  Within each sample we also give  the differences in deviance
 $d\mathcal{L}$ (Eq.\,\ref{e:diffitness}) between each model and the model with a single
 isotropic maxwellian, which gives an indication of their relative merits.}
\end{table*}

The isotropic maxwellian velocity distribution has a major advantage
in enabling us to compute three out of four integrals in
Eq.\,\ref{e:jointmaxwel} analytically. 
However, once the pulsar has moved away from the galactic plane,
we have more information, that we will put to use in this Section:
the pulsar velocity must be directed away from its place of birth,
which for sufficiently large $|z|$ implies that $v_z>0$ when $z>0$ and
$v_z<0$ when $z<0$. For these pulsars we assume an intrinsic distribution for
the velocity which is an isotropic distribrution from which the
velocities towards the galactic plane have been removed:
and refer to this distribution as semi-isotropic.

To quantify `sufficiently large' we show the nominal values of
distance to the galactic plane $z'=D'\sin b$ in Fig.\,\ref{f:lzvz},
together with a band indicating the scale height of O stars, as
a proxy for the place of birth of pulsars. Five pulsars in our list
of 28 are moving towards the galactic plane. Two of these,
PSR\, B0329+54 (\#4 in our master list) and 
PSR\,J0538+2817 (\#7) are within the region where pulsars are
born, and thus may well be moving towards the plane. 
PSR\,J2144$-$3933 (\#26) is the oldest pulsar in our sample,
and may well be a returning pulsar. PSR\,B0818$-$13 (\#11)
and PSR\,B1237+25 (\#15) are too young -- assuming their
characteristic age is indicative of their real age -- to have reversed
motion, and their motion towards the galactic plane must be 
apparent. We may write $v_z$ as (see Fig.\,\ref{f:syscoor})
\begin{equation}
v_z = D\mu_b\cos b + v_r\sin b
\label{e:vzero}\end{equation}
hence a pulsar is moving away from the plane if
\begin{equation}
zv_z>0 \,\mathrm{if}\, v_r>{-\mu_b\cos b\over \varpi\sin b}
\end{equation}
Entering the nominal values $\varpi'$ and $\mu'_b$, we obtain
$v_r>120$\, km/s (\#11) and $v_r>12$\,km/s (\#15), indicating
that these pulsars may well be moving as expected: away from the plane.

In computing for the case of semi-isotropic maxwellians, we choose axes
parallel to the (local) direction of right ascension and declination,
and along the line of sight, and write the spatial velocity as
\begin{equation}
\vec{v}=(v_\alpha,v_\delta,v_r)= (v\sin\xi_1\cos\xi_2, v\sin\xi_1\sin\xi_2,v\cos\xi_1)
\label{e:vr}\end{equation}
where $$ 0\leq\xi_1\leq\pi; \qquad 0\leq\xi_2\leq2\pi$$
To determine which velocities lead to $v_z$ away from the galactic
plane, we first convert the velocities to galactic coordinates using
Eqs.\,\ref{e:mul}, \ref{e:mub}:
\begin{equation}
(v_l,v_b,v_r)=
(v\sin\xi_1\cos(\xi_2-\phi),v\sin\xi_1\sin(\xi_2-\phi),v\cos\xi_1)
\label{e:vbe}\end{equation}
where $\phi$ is given by Eq.\,\ref{e:phi}. 
Entering $v_b$ and $v_r$ from Eq.\,\ref{e:vbe} into Eq.\,\ref{e:vzero}
we obtain
\begin{equation}
v_z=v[\sin\xi_1\sin(\xi_2-\phi)\cos b + \cos\xi_1\sin b]
\label{e:vzcorrect}\end{equation}
Note that the sign of $v_z$ {\em does not depend on the speed} $v$.
The condition $v_z>0$ if $b>0$ and  $v_z<0$ if $b<0$ may be written
\begin{equation}
\sin(\xi_2-\phi) > {-\tan b\over\tan\xi_1}
\label{e:condition}\end{equation}
We rewrite the joint probability of Eq.\,\ref{e:definejm} for the
semi-isotropic case as
$$
P_\mathrm{sim}(\varpi',\mu'_{\alpha*},\mu'_\delta,D,v,\xi_1,\xi_2)=0\qquad
\mathrm{if}\quad zv_z<0
\phantom{oliebololiebololi} $$
$$ P_\mathrm{sim}(\varpi',\mu'_{\alpha*},\mu'_\delta,D,v,\xi_1,\xi_2)=\mathcal{C}
\exp\left[-\,{(1/D-\varpi')^2\over2{\sigma_\varpi}^2}\right] \phantom{oliebololiebol}$$
\begin{eqnarray}
&\times&\exp\left[-\,{(\mu_{\alpha*,G}(D)+v\sin\xi_1\cos\xi_2/D
-\mu'_{\alpha*})^2\over2{\sigma_\alpha}^2}\right] 
\nonumber \\
&\times&\exp\left[-\,{(\mu_{\delta,G}(D)+v\sin\xi_1\sin\xi_2/D
-\mu'_{\delta})^2\over2{\sigma_\delta}^2}\right]   \nonumber \\
\phantom{olie}& \times & 
f_D(D)\sin\xi_1\,2\sqrt{\frac{2}{\pi}}{v^2\over\sigma^3}\exp\left[-\,{v^2\over2\sigma^2}\right]
\quad \mathrm{if}\quad zv_z>0
\label{e:definejmsi}\end{eqnarray}
where $\mathcal{C}$ is defined with Eq.\,\ref{e:fin_maxw}, and a
factor 2 is added to normalize the semi-maxwellian.
The likelihood for the semi-isotropic maxwellian follows:
\begin{equation}
L_\mathrm{sim}(\sigma) =
\int_o^{D_\mathrm{max}} \int_0^{2\pi} \int_0^\pi\int_0^\infty
 P_\mathrm{sim}dv d\xi_1 d\xi_2 dD
\label{e:jointmaxwelsi}\end{equation}
Eq.\,\ref{e:vzcorrect} shows that the condition that $v_z$ is in the
correct direction is determined by the angles $\xi_1$ and
$\xi_2$ and does not depend on $v$, and this allows the
integral in Eq.\,\ref{e:jointmaxwelsi} over the velocity to be done analytically.
The integrals over the angles and distance are done numerically. Details are
given in Appendix\,\ref{a:num_int}.

\section{The mixed model\label{s:mixed}}

\begin{figure}
\centerline{\includegraphics[width=0.8\columnwidth]{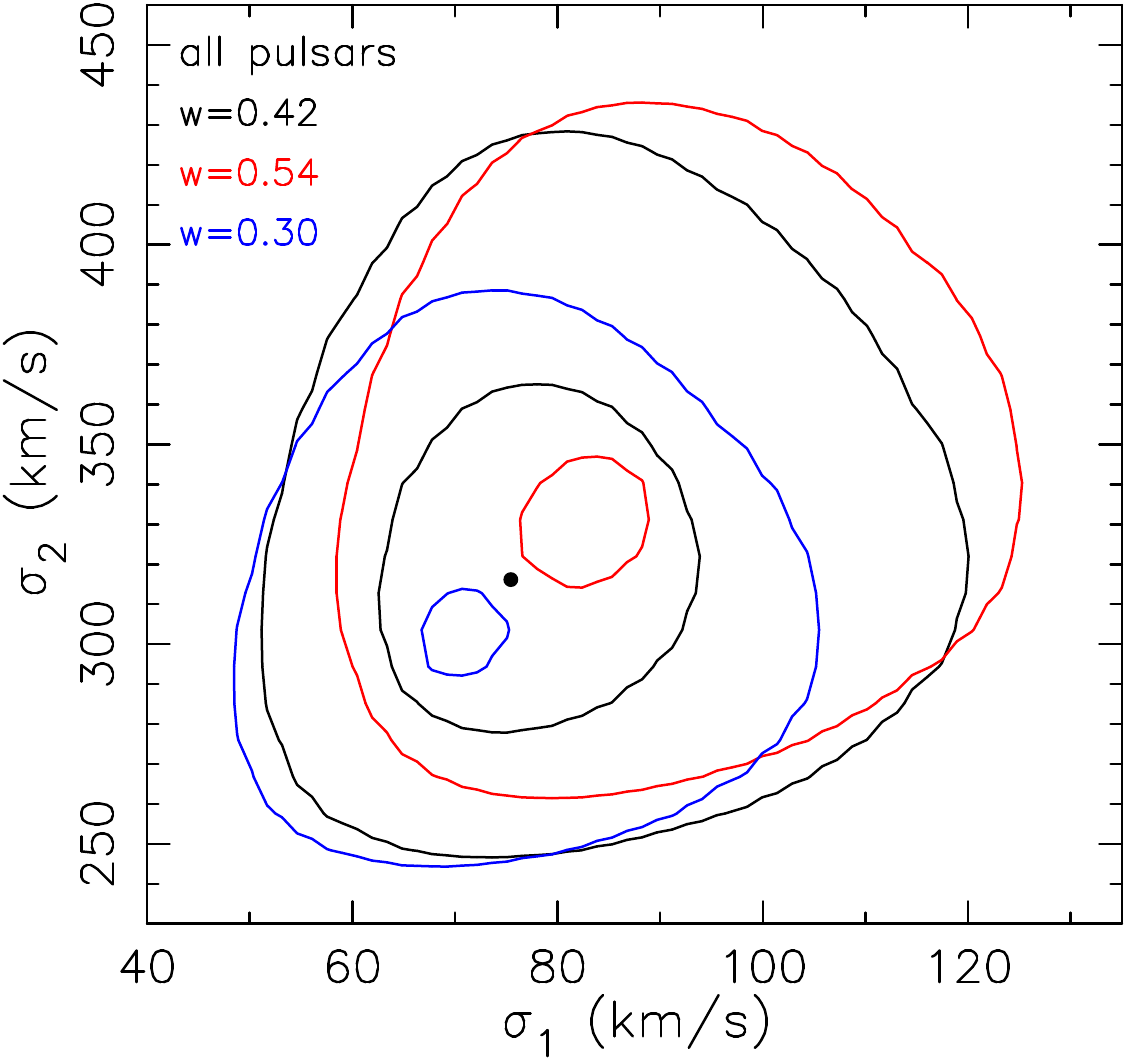}}
\centerline{\includegraphics[width=0.8\columnwidth]{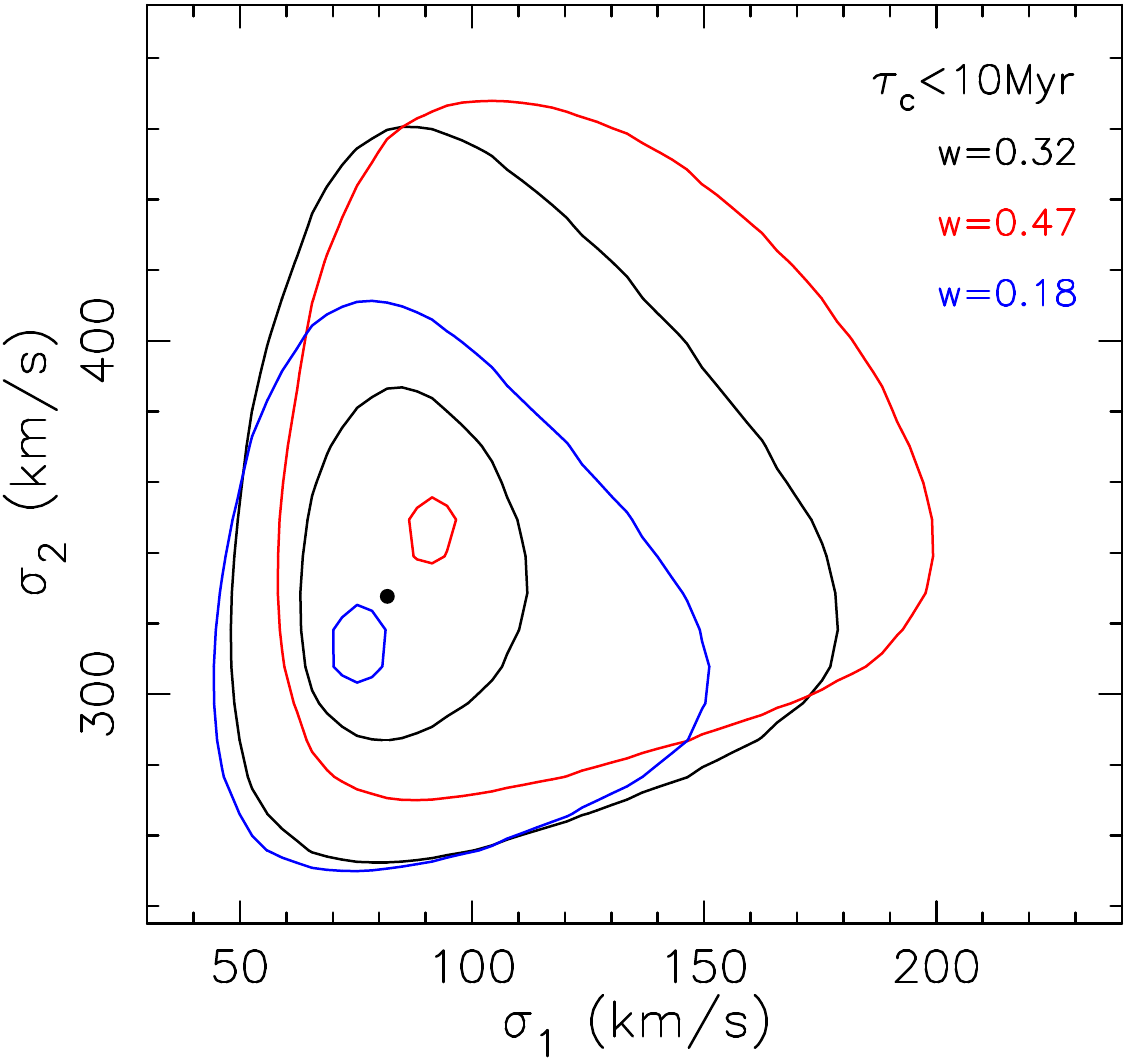}}
\caption{As Figs.\,\ref{f:twomaxw}, now for the
  mixed model.  $\vec{\sigma}_\mathrm{opt}$ for all pulsars and for 
  the youngest pulsars are listed in Table\,\ref{t:results}.
 \label{f:mixed}}
\end{figure}

In our mixed model we assume that
the pulsars in the grey band in Fig.\ref{f:lzvz} (\#4,5,7,9,19) and
the oldest pulsar (\#26) are  drawn from an isotropic velocity
distribution, whereas all others are drawn from a semi-isotropic
distribution, in which the velocities towards the galactic plane are
excluded. The distribution parameter $\sigma$ for the 
semi-isotropic distribution is equal to the $\sigma$ for the
isotropic distribution.
In analogy with Eqs.\,\ref{e:deviance} we define the deviance for the
mixed model as
\begin{equation}
\mathcal{L}_{\mathrm{mixed}} (\sigma) =
 -2\left(\sum_i\ln L_{\mathrm{sim},i} (\sigma)
 +\sum_j\ln L_{\mathrm{maxw},j} (\sigma)\right)
\label{e:max_liksi}
\end{equation}
where the sums over $i$ and over $j$ are for the pulsars whose
velocity is drawn from a semi-isotropic distribution and an isotropic
distribution, respectively.  The best value for $\sigma$ is the value
for which Eq.\,\ref{e:max_liksi} reaches its minimum, and its range is
determined from $\Delta\mathcal{L}=1$. The results are given in
Table\,\ref{t:results} and Fig.\,\ref{f:singlemax}, and are not very
different from those for the single isotropic maxwellian, both for
sample A of all pulsars, and for sample Y for the youngest pulsars.
For the $d\mathcal{L}$ value it is seen that the mixed model is a
significant improvement on the isotropic model. We return to this
below, for the more interesting case of two maxwellians.

In a more realistic model the semi-isotropic distribution is composed of two
semi-maxwellians, with the same distribution parameters $\vec\sigma$
as the two isotropic maxwellians that compose the isotropic
distribution.

In analogy with Eqs.\,\ref{eq:2maxw} we now have
\begin{equation}
L_{2\mathrm{sim}} (\vec\sigma) = w L_\mathrm{sim} (\sigma_1) + (1-w) L_\mathrm{sim} (\sigma_2)
\label{e:2maxwsi}
\end{equation}
and in analogy with Eqs.\,\ref{e:max_liksi} 
\begin{equation}
\mathcal{L}_{2\mathrm{mixed}} (\vec\sigma) =
 -2\left(\sum_i\ln L_{2\mathrm{sim},i} (\vec\sigma)
 +\sum_j\ln L_{2\mathrm{maxw},j} (\vec\sigma)\right)
\label{e:2max_liksi}
\end{equation}
where the sums over $i$  and over $j$ are for the pulsars whose
velocity is drawn from a semi-isotropic distribution and an isotropic
distribution, respectively. 
We use the subroutine {\tt AMOEBA} of Press et
al.\ (1986) \nocite{press86} to obtain the optimal values of $w$, $\sigma_1$ and
$\sigma_2$ for which $\mathcal{L}_{2\mathrm{mixed}}$ has its minimum,
and $\Delta\mathcal{L}=1$ for the range of these parameters.
The results are given in Table\,\ref{t:results} and Fig.\,\ref{f:mixed}.

The best values and the ranges for $\sigma_1$, $\sigma_2$
and $w$ for the semi-isotropic model are not significantly different
from those of the isotropic model.  Contour plots in the
$\sigma_1$-$\sigma_2$ planes also are not significantly different from
those for the model with two isotropic maxwellians shown in
Figs.\,\ref{f:twomaxw}. 

The factor 2 in (the last line of) Eq.\,\ref{e:definejmsi} ensures that the semi-maxwellian
is normalized to unity. As remarked above, a constant multiplicative factor for any likelihood
drops out in Eqs.\,\ref{e:dfitness}, and thus does not affect the best solution and its
range(s) within one model.
However, to compare between models one must use the same
normalizations of the separate distributions between
the different models, and this requires the factor 2 in Eq.\,\ref{e:definejmsi}.
The $d\mathcal{L}$ values listed in Table\,\ref{t:results} show that
the mixed model is a highly significant improvement above
the isotropic maxwellian model, for the full sample A, and that it is still
significant  for sample Y of young pulsars. 

It is interesting to look at this is some more detail. Suppose for the
moment that the contributions to the integral of
Eq.\,\ref{e:jointmaxwel} are zero for $v_z$ velocities towards the
plane, then the only difference between
$L_{2\mathrm{maxw}}(\vec\sigma)$ and $L_{2\mathrm{mixed}}(\vec\sigma)$ is the
multiplicative factor 2 in Eq.\,\ref{e:definejmsi}.  In sample\,A for
all pulsars, this affects only the 22 pulsars for which a semi-maxwellian
applies, and leads to an added term in Eq.\,\ref{e:2max_liksi} equal to
$-2\times22\times\ln2\simeq-30.5$.  In sample\,Y 14 of the young
pulsars are affected, leading to an added term
$-2\times14\times\ln2\simeq-19.4$. The actual differences $d\mathcal{L}$
in deviance between the mixed models and purely isotropic models
are smaller than this, which indicates that $v_z$ velocities towards
the plane {\em do} contribute to the integral of
Eq.\,\ref{e:jointmaxwel}, also for pulsars for which such velocities
are not expected. This implies that the isotropic model overestimates
the likelihoods for these pulsars. PSR\,B0818$-$13 (\#11) is a case 
in point: its apparent $v'_z$ velocity is towards the plane
(Fig.\,\ref{f:lzvz}), and thus $v_z$ velocities towards the plane
may be expected to contribute noticeably to integral
Eq.\,\ref{e:jointmaxwel}.

\begin{figure}
\centerline{\includegraphics[width=\columnwidth]{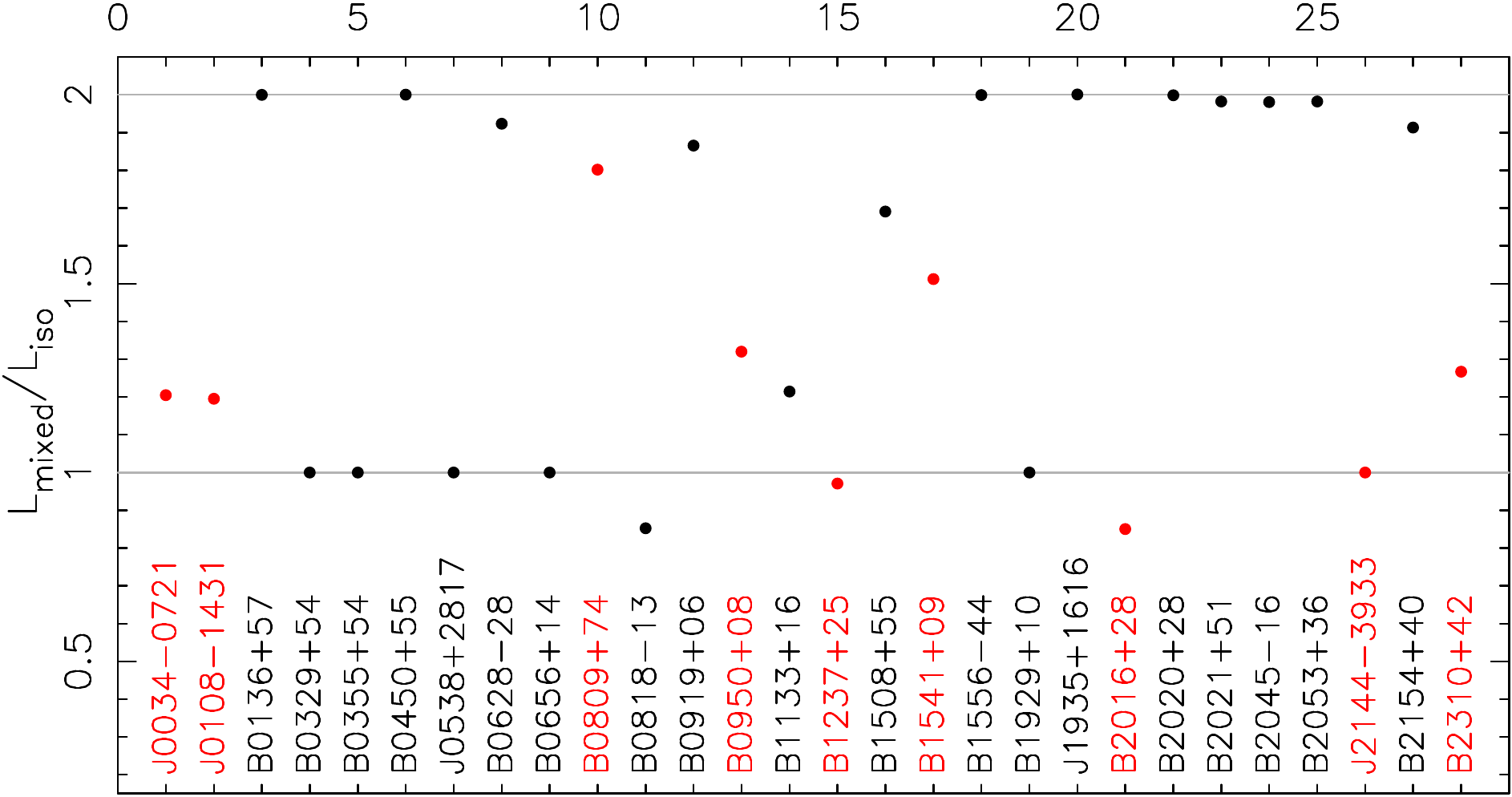}}
\caption{For each pulsar the ratio of the likelhoods in the mixed and
  isotropic models is shown. The red colour indicates old pulsars,
  with $\tau_c>10$\,Myr. To illustrate the pure effect of the
  normalization of the velocity distribution, we use the same parameters
 $\vec\sigma = $($76$km/s, $318$km/s, 0.32) for both
 likelihoods. Use of $\vec\sigma_\mathrm{opt}$ for each model separately gives
 rise to small shifts.
 \label{f:ratiolik}}
\end{figure}

In Fig.\,\ref{f:ratiolik} we show the ratio of the likelihoods for the
mixed and isotropic two-maxwellian model for each pulsar separately.
The six pulsars whose velocities are drawn from an isotropic velocity
distribution also in the mixed model by definition have a ratio of
one of the likelihoods for the mixed and isotropic two-maxwellian
model. The eleven pulsars with ratios closest to the maximum possible,
$1.8<L_\mathrm{mixed}/L_\mathrm{iso}<2$ say, are all young. For these
pulsars, almost all velocities contributing to $L_i$ in the isotropic model
contribute also in the mixed model. For 11 pulsars (sample A) or 3
pulsars (sample Y) the velocity range that contributes to $L_i$ is
restricted by the condition that $v_z$ be away from the galactic
plane, as shown by the difference of their
$L_\mathrm{mixed}/L_\mathrm{iso}$ from the normalization factor 2. 

\section{The distribution of longitudinal velocities\label{s:gauss}}

\begin{figure}
\centerline{\includegraphics[width=\columnwidth]{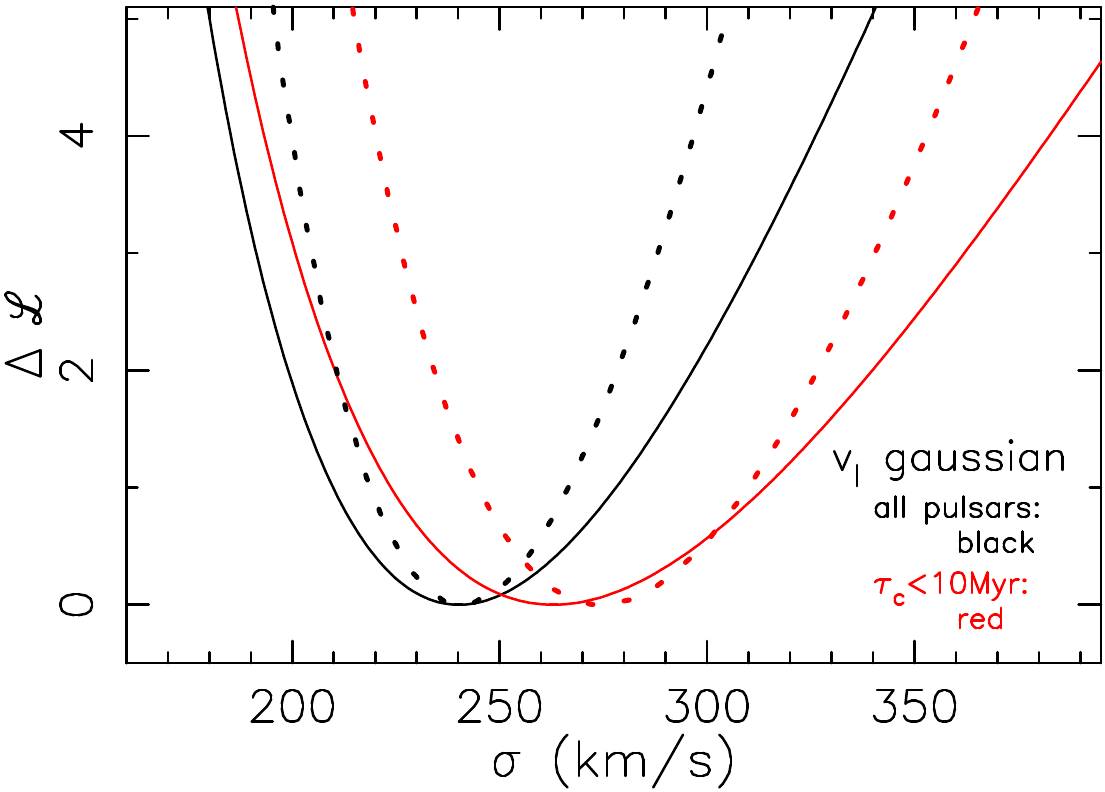}}
\caption{Variation of $\mathcal{L}$ with $\sigma$ when only
 measurements of the parallax $\varpi'$ and of the proper motion
 $\mu'_{l*}$ in the direction of galactic longitude are used (solid
 lines). For comparison the results for the mixed model, that uses
 parallaxes and both proper motions $\mu_{\alpha*},\mu_\delta$ are also
 shown (dotted lines).
 \label{f:gauss}}
\end{figure}

\begin{table*}
\caption{Comparison of the results of our best model with those
  obtained in some earlier studies\label{t:previous}}
\center{
\begin{tabular}{|l|rc|rcrcrc|}
\hline
 & \multicolumn{2}{c|}{single maxwellian} &
\multicolumn{6}{c|}{two maxwellians} \\
\hline
 & $\sigma$ & range & $\sigma_1$ & range & $\sigma_2$ & range & w &
 range \\
reference  & \multicolumn{2}{c|}{(km/s)} & \multicolumn{2}{c}{(km/s)} &
 \multicolumn{2}{c}{(km/s)} & \multicolumn{2}{c|}{(\%)} \\
\hline
Arzoumanian et al.\ (2002)$^a$ & 290 & 260-320 & 90 & 75-110 & 500 &
350-750 & 40 & 20-60  \\
Brisken et al.\ (2003a) & & & 99 & & 294 & & 20 & \\
Hobbs et al.\ (2005) & 265 & 239- 291 & & & & & & \\
Faucher-Gigu\`ere \&\ Kaspi (2006) & 290 & 260-320 & 160 & 130-180 & 
780 & 640-930 & 90 & 87-100 \\
mixed model sample A     & 239 & 219-267 & 75 & 60-95 &  316 & 276-368
& 42 & 30-52 \\
\hline
\end{tabular}}
\tablefoot{$^a$Arzoumanian et al.\ (2002)  fit
  gaussians; comparison of their Eq.\,1 with our
  Eq.\,\ref{e:twomaxw} shows that these are components of maxwellians.
  Thus, their $\sigma$ values may be compared directly
  with those in the other papers, contrary to the statement by Brisken et al.\ (2003a, below
  their Eq.\,3).}
\end{table*}

For comparison with earlier studies we also determine the model
parameter $\sigma$ by only using the measurements of the
parallax and the measurements $\mu'_{l*}$
of the proper motion in the direction of galactic longitude. 
For this we choose the coordinates in the directions of
galactic longitude and latitude, and radial. We
rewrite Eqs.\,\ref{e:definejm} and \ref{e:jointmaxwel} as
$$
P_\mathrm{gauss}(\varpi',\mu'_{l*},D,v_l,v_b,v_r)=\mathcal{C}_l
f_D(D)G(v_l,\sigma)G(v_b,\sigma)G(v_r,\sigma)\phantom{olieb}$$
\vspace*{-0.5cm}
\begin{eqnarray}
\phantom{oliebolbakker}&\times&\exp\left[-\,{(1/D-\varpi')^2\over2{\sigma_\varpi}^2}\right] \nonumber \\
&\times&\exp\left[-\,{(\mu_{l*,G}(D)+v_l/D
-\mu'_{l*})^2\over2{\sigma_l}^2}\right] 
\label{e:jointl}\end{eqnarray}
where
$$ \mathcal{C}_l \equiv
\left[2\pi\sigma_\varpi\sigma_l\int_0^{D_\mathrm{max}} f_D(D)
  dD\right]^{-1} $$ 
and
\begin{equation}
L_\mathrm{gauss}(\sigma) =
\int_o^{D_\mathrm{max}}\int_{-\infty}^\infty\int_{-\infty}^\infty\int_{-\infty}^\infty
 P_\mathrm{gauss}dv_l dv_b dv_r dD
\label{e:gaussb}\end{equation}
$\mu_{l*}$ and its error $\sigma_\mu$ are obtained from $\mu_{\alpha*}$, $\mu_\delta$
and their errors with Eq.\,\ref{e:mul}. 
The integrals over $v_b$ and $v_r$ are decoupled from the other integrals, and
equal to 1. Eq.\,\ref{e:gaussb} is rewritten:
\begin{equation}
L_\mathrm{gauss}(\sigma) =\mathcal{C}_l\int_o^{D_\mathrm{max}}f_D(D)
\exp\left[-\,{(1/D-\varpi')^2\over2{\sigma_\varpi}^2}\right] \mathcal{I}_l\,dD
\label{e:likgauss}\end{equation}
where
$$\mathcal{I}_l  \equiv \left(1+{\sigma^2\over D^2{\sigma_l}^2}\right)^{-1/2}
\exp\left[-{1\over2}{(D\,\mu_{l*,G}-D\,\mu'_{l*})^2\over
  \sigma^2+D^2{\sigma_l}^2}\right]$$
Note that in this case, there is no difference
between the isotropic and mixed model, because $v_z$ does not affect
$v_l$.  We compute the deviance (Eq.\,\ref{e:deviance}) with
Eq.\,\ref{e:likgauss} , to determine the values $\sigma_\mathrm{opt}$
for which the deviance reaches its minimum, and their range from
$\Delta\mathcal{L}=1$. The results are listed in Table\,\ref{t:results} and
shown in Fig.\,\ref{f:gauss}. Interestingly, \psrbad\ is not an
outlier in $v_l$: its proper motion is almost completely in the
direction of galactic latitude (see Fig.\,\ref{f:observed}). For sample
A (all pulsars), $\sigma_\mathrm{opt}$ is the same as for the
isotropic or semi-isotropic single maxwellian; for sample Y
(youngest pulsars) it is marginally lower. The limitation to
only one component of the proper motion leads to a reduced accuracy of
$\sigma_\mathrm{opt}$, as expected.
As a consequence the superposition of two gaussians (i.e.\ components
of two maxwellians in the direction of galactic longitude)
does not improve significantly over the single maxwellian
description ($\vec\sigma=109\,\mathrm{km/s},277\,\mathrm{km/s},0.27$,
$d\mathcal{L}=1$). 

\section{Conclusions and discussion\label{s:discussion}}

Previous work derived the velocity distribution of pulsars from the
observed distances and proper motions, and then compared this
distribution with model distributions.  This reduces the information
present in the observations, complicates error propagation, and has
lead to wrong likelihood definitions.  The uncertainties in the proper
motions determined from timing are two to three orders of magnitude
larger than those of the proper motions in our master list, that are
determined from VLBI.  The larger number of such proper motions (less
than one order of magnitude) does not make up for their larger
uncertainties, so that inclusion of these proper motions does not significantly
improve the analysis.  The use of distances determined from dispersion
measures further complicates the analysis, because the related
distance uncertainties are dominated by systematic effects, and cannot
be described with a gaussian, even in approximation.

Our approach is more reliable because we a) derive predictions for the
observed parameters (parallax and proper motion) from the model, and
compare these directly with the relevant measurements, b) only use
VLBI determinations from after 2000 of both parallax and proper motion, whose
uncertainties are well described with gaussians, and c) include the
intrinsic galactic distribution of pulsars (as expressed in $f_D(D)$,
Eq.\,\ref{e:fd}). Our mixed model furthermore takes into account that
velocity component $v_z$ perpendicular to the galactic plane of a
young pulsar well away from that plane must be in the direction away
from the plane.

Applying this to the pulsars in our master list, we find that the
description of the velocity distribution of the pulsars with two
maxwellians improves significantly on the description with a single
maxwellian. Our model describing $v_l$ with a single gaussian gives a
similar value for $\sigma$ as the (mixed or isotropic) single
maxwellian, as expected for an isotropic velocity distribution.
Comparison with earlier results, compiled in Table\,\ref{t:previous},
shows that our more accurate method leads to more accurately
determined model parameters.  We show in Fig.\,\ref{f:observed} that
our best solution corresponds well with the observed distribution of
$v_\perp$. One would be tempted to conclude that our whole
analysis apparatus can be replaced with a straightforward fit
of the cumulative $v_\perp$ according to the model to the
observed cumulative data for $v'_\perp$! The reasons for
the succes of the simpler method are the relatively small
errors in the parallax, which limit the importance of $f_D(D)$,
and the smallness of the correction for galactic rotation
with respect to the observed proper motions: 
$\mu_{\alpha*,G}\ll\mu'_{\alpha*}$ and $\mu_{\delta,G}\ll\mu'_\delta$
(Fig.\,\ref{f:observed}). Indeed, ignoring the corrections for
galactic rotation hardly affects the results (Verbunt \&\ Cator 2017).
\nocite{verbuntcator17}
Corrections for galactic motion matter only for distances much larger 
than those of the pulsars in our master list. 

With the exception of Brisken et al.\ (2003a), who do not give error
estimates, all previous authors find significantly higher velocities
for the high-velocity component than we do. The compilation in
Table\,\ref{t:previous} illustrates that the fraction of pulsars in
the high-velocity component (i.e.\ $1-w$) is inversely related to the
characteristic velocity of that component. A small number of
erroneously very high velocities leads to a high value of $\sigma_2$.
Because the combination of $\sigma_2>500$\,km/s with a low value of
$w$, i.e. high $1-w$, would lead to a much higher fraction of pulsars with
$v_\perp>370$\,km/s, say, than observed, the high value of $\sigma_2$ forces
a low value of $w$.  We suggest that the higher velocities derived
by previous authors are affected by the inclusion of unreliable
distances determined from dispersion measures.  In the case of
Arzoumanian et al.\ (2002) we note that all parallaxes are from before
2000, i.e.\ not corrected for differential ionospheric refraction. As
Hartman (1997) \nocite{hartman97} has shown, underestimating velocity errors leads to
overestimating velocities.

The analysis by Hobbs et al.\ (2005) is based on the nominal
velocities $v'_\perp=\mu'_\perp/\varpi'$, and does not take
into account the large errors in both distances and proper motions
of their sample. These errors blur the intrinsic distribution. 
We suggest that this prevents Hobbs et al.\ from recognising the
presence of low velocities, and from
recovering a bimodal velocity distribution in their analysis.
The best model with two velocity components by  Faucher-Gigu\`ere \&\ Kaspi
(2006) allows $w=1$, i.e.\ the second component is not significant. 
Our analysis in Sect.\,\ref{s:gauss} suggests that this is due to their small sample
size (34 pulsars, of which only 8 have a measured parallax). 

\begin{figure}
\centerline{\includegraphics[width=\columnwidth]{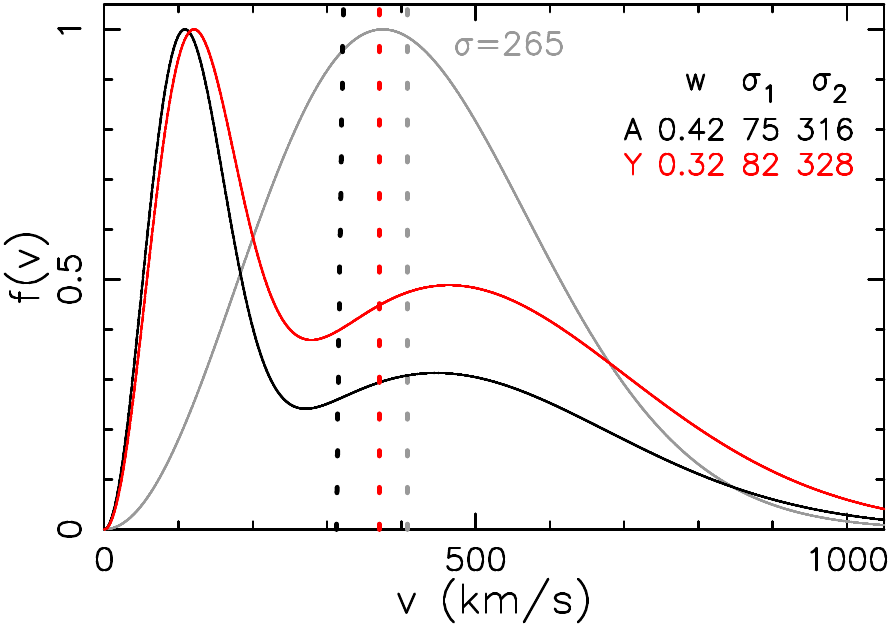}}
 \vspace*{0.2cm}

\centerline{\includegraphics[width=\columnwidth]{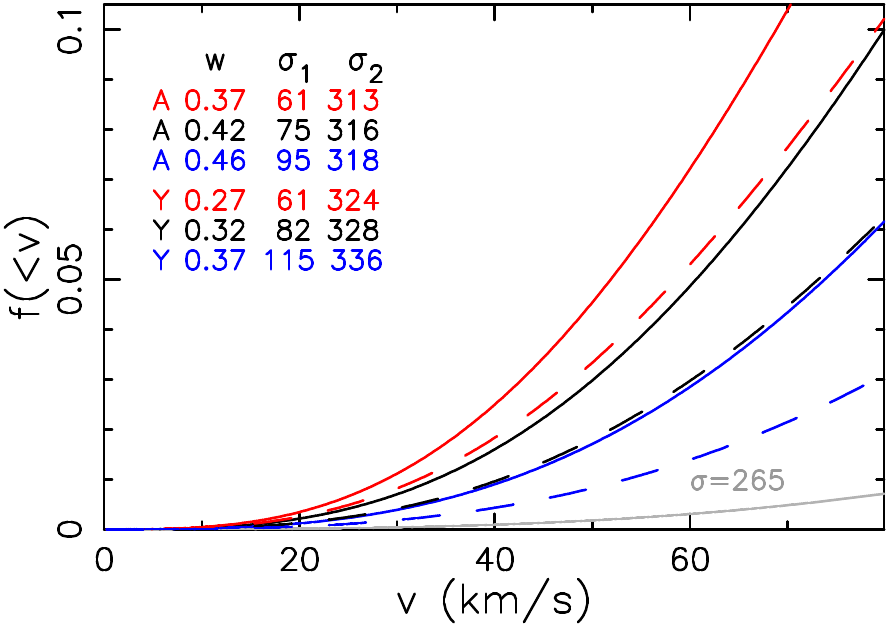}}
\caption{Top: our best velocity distribution for all pulsars and for
  the youngest pulsars, together with a single maxwellian. The
  vertical dotted lines indicates the median velocities: 313, 370
  and 408 km/s. Below:
  fraction $f(<v)$ of pulsars with velocity less than $v$, for
  the best mixed model for all pulsars (black), and for the lowest
  and highest value in the  range of $\sigma_1$ (red,blue). Solid lines: all pulsars; dashed
  lines: pulsars with $\tau_c<10$\,Myr. In grey we show the fraction
  for a single maxwellian (Hobbs et a.\ 2005).
  \label{f:survive}}
\end{figure}

Our results imply that the velocity contrast between the
low- and high-velocity components is a factor 3 to 6, and that 30 to
50\%\ of the pulsars arise from the low-velocity component.
It has been suggested (Podsiadlowksi et al.\ 2004) \nocite{podsiadlowski04}
that pulsars formed from small iron cores or via electron capture would have a lower
kick velocity than those formed from higher-mass core collapse,
which may lead to a bimodal velocity distribution of pulsars.
The existence of a class of neutron stars with low birth velocity has
been derived from the properties of Be X-ray binaries (Pfahl et al.\
2002) and the properties of millisecond pulsars binaries (Van den
Heuvel 2004).\nocite{pfahl02}\nocite{vandenheuvel04}

The fact that some pulsars are born in binaries and others from single
stars will also affect the velocity distribution of single pulsars.
Whether the observed bimodal velocity distribution reflects these different
origins can be investigated in a population synthesis.

One of the goals of our work was to determine the fraction of pulsars
with velocities small enough to remain bound to a globular cluster, or
in a binary. In Fig.\,\ref{f:survive} we show the fraction of pulsars
with velocity less than $v$ as a function of $v$. For a typical escape
velocity of a globular cluster, 60 km/s say, it is seen that this fraction is about
5\%\ in our best model (mixed, for sample A). It varies from about
3\%\ to about 7.5\%\ in the range of $\sigma_1$. At these low
velocities, the fraction of pulsars is dominated completely by the
low-velocity component, and  therefore varies linearly
with $w$ for fixed $\sigma_1$ and $\sigma_2$.

Finally, we mention two reasons why the determination of pulsar velocities
from a local sample may lead to an underestimate of the average velocity.
The first one is galactic drift: motion in the galactic gravitational potential
leads to  reduction of the velocity of a pulsar that moves away from
the center of the galaxy, and an increase if it moves towards the center.
Thus if  pulsars with an origin closer to the galactic center
contribute more to the locally observed sample than pulsars with an origin
further out, the locally measured velocity distribution underestimates
the distribution at birth (Hansen \&\ Phinney 1997). \nocite{HansenPhinney1997}
The second reason
is related to the velocity perpendicular to the plane: pulsars with a
high $|v_z|$ move further from the plane, and thus must have a higher
luminosity to be detected. In a flux-limited sample this leads to
an over-representation of the low-velocity pulsars.
These effects can be studied best in a population synthesis
that takes these and other selection effects into account.
Since such a synthesis involves also a larger number of parameters,
a first step would be the measurement of more pulsar distances and
proper motions.

\begin{acknowledgements}
We thank Gijs Nelemans for useful discussions.
\end{acknowledgements}

\begin{appendix}
\section{Transformations of equatorial to galactic coordinates\label{s:conversions}}
For the convenience of the reader we summarise the equations
for coordinate transformations that we use.
Lane (1979) \nocite{lan79} gives (two of the three) equations for
conversion from galactic to equatorial for B1950.0. He notes that the
equatorial coordinates of the galactic pole $\alpha_\mathrm{GP}$,
$\delta_\mathrm{GP}$ and the galactic longitude $l_\Omega$ of the node
where the galactic plane ($b=0$) crosses the equator, define the
coordinate transformation and thus also the equatorial coordinates of
the centre $l=b=0$. Note that this centre does not coincide exactly
with the actual centre of the galaxy (e.g.\ as defined by Sgr\,A$^*$).
We give all three equations, rewriting them slightly to show
explicitly the role of $\alpha_\mathrm{GP}$, $\delta_\mathrm{GP}$, and
$l_\Omega$.  The coordinate transformation is composed of three
rotations: one around the galactic $z$-axis to bring the galactic
centre to the node (this replaces $l$ with $l-l_\Omega$), one
around the equatorial $z$-axis to bring the spring node to the node
(this replaces $\alpha$ with
$\alpha-\alpha_\Omega=\alpha-(\alpha_\mathrm{GP}+{\pi\over2}$), and
finally around the now common $x$-axis over an angle
${\pi\over2}-\delta_\mathrm{GP}$ to align the galactic pole with the
equatorial pole. The resulting equations are (see also Lane 1979).
\begin{eqnarray}
\label{e:eqtogala} \cos(\alpha-\alpha_\mathrm{GP}-{\pi\over2})\cos\delta
&=&\cos(l-l_\Omega)\cos b \\
\sin(\alpha-\alpha_\mathrm{GP}-{\pi\over2})\cos\delta&=&
\cos({\pi\over2}-\delta_\mathrm{GP})\sin(l-l_\Omega)\cos b
\nonumber  \\
& & \label{e:eqtogalb} -\sin({\pi\over2}-\delta_\mathrm{GP})\sin b \\
\sin \delta &=&
\sin({\pi\over2}-\delta_\mathrm{GP})\sin(l-l_\Omega)\cos b 
\nonumber \\ 
& &\label{e:eqtogalc} +\cos({\pi\over2}-\delta_\mathrm{GP})\sin b
\end{eqnarray}

To find the equatorial coordinates $\alpha_\mathrm{GC}$,
$\delta_\mathrm{GC}$ for the centre of the coordinate
system, we enter $l=b=0$ and combine
eqs.\,\ref{e:eqtogala},\ref{e:eqtogalb} to find:
\begin{eqnarray}
\tan (\alpha_\mathrm{GC}-\alpha_\mathrm{GP}-{\pi\over2}) & = &\label{e:eqtogald}
{ \cos({\pi\over2}-\delta_\mathrm{GP})\sin(-l_\Omega) \over \cos(-l_\Omega)}\\
\sin \delta_\mathrm{GC} &=&\label{e:eqtogale}
\sin({\pi\over2}-\delta_\mathrm{GP})\sin(-l_\Omega) 
\end{eqnarray}
Perryman et al.\ (1997) give the pole and node longitude for J2000.0 as
\begin{equation}
\alpha_\mathrm{GP} = 192\fdg85948, \quad
\delta_\mathrm{GP} = 27\fdg12825, \quad
l_\Omega = 32\fdg93192
\label{e:eq2000a}\end{equation}
and with Eqs.\,\ref{e:eqtogala},\ref{e:eqtogalb},\ref{e:eqtogalc},
these define the coordinate transformation for J2000.0 in the
ICRS system. Entering these values in Eqs.\,\ref{e:eqtogald},
\ref{e:eqtogale} we find
\begin{equation}
\alpha_\mathrm{GC}= 266\fdg40500, \quad 
\delta_\mathrm{GC} = -28\fdg93617
\label{e:eq2000b}\end{equation}
For later reference we combine Eqs.\,\ref{e:eqtogala},
\ref{e:eqtogalb} for the galactic center $l=b=0$ into
\begin{equation}
\tan(-l_\Omega) ={\sin (\alpha_\mathrm{GC}-\alpha_\mathrm{GP}-{\pi\over2})
/\cos({\pi\over2}-\delta_\mathrm{GP})\over
\cos (\alpha_\mathrm{GC}-\alpha_\mathrm{GP}-{\pi\over2})}
\label{e:eqtogalf} \end{equation}
and note that entering the coordinates for pole and centre from 
Eqs.\,\ref{e:eq2000a}, \ref{e:eq2000b} in Eq.\,\ref{e:eqtogalf}
we re-obtain $l_\Omega$ correctly. 

The next step is to determine the transformation of the
proper motions. This is done by Smart (1938, chapter 1.41)\nocite{smart1938}, who notes that
it corresponds to a rotation over an angle $\phi$ between
the local directions of the lines of constant $l$ and 
constant $\alpha$, or equivalently between the lines of constant $b$ and
constant $\delta$. With the notation $\mu_{l*}\equiv\mu_l\cos b$ and
$\mu_{\alpha*}\equiv\mu_\alpha\cos\delta$ we write Smart's Eqs.\,4,5 as
\begin{eqnarray}
\mu_{l*}&=&\phantom{-}\mu_{\alpha*}\cos\phi + \mu_\delta\sin\phi \label{e:mul}\\
\mu_b & = & -\mu_{\alpha*}\sin\phi  + \mu_\delta\cos\phi\label{e:mub}
\end{eqnarray}
From spherical trigonometry the angle $\phi$ is given by
\begin{equation}
\tan \phi = {\sin(\alpha-\alpha_\mathrm{GP})\over
\cos\delta\tan\delta_\mathrm{GP}-\sin\delta\cos(\alpha-\alpha_\mathrm{GP})}
\label{e:phi}\end{equation}
(Smart 1938, Eq.3). The angle $\phi$ may also be found by taking the
time derivative of the equation defining the transformation
equatorial coordinates to galactic latitude (cf.\ Lane 1979)
\begin{eqnarray}
\sin b & = & \sin \delta\cos({\pi\over2}-\delta_\mathrm{GP}) \nonumber \\
& & -\sin({\pi\over2}-\delta_\mathrm{GP})\sin(\alpha-\alpha_\mathrm{GP}-{\pi\over2})
\cos\delta
\end{eqnarray} 
and equating the result to Eq.\,\ref{e:mub}.

For galactic to equatorial we may write analogously to Eqs.\ref{e:mul}
and \ref{e:mub}:
\begin{eqnarray}
\mu_{\alpha*}&=&\phantom{-}\mu_{l*}\cos\phi_2 + \mu_b\sin\phi_2 \label{e:mua}\\
\mu_\delta & = & -\mu_{l*}\sin\phi_2  + \mu_b\cos\phi_2\label{e:mud}
\end{eqnarray}
We equate the time  derivative of Eq.\,\ref{e:eqtogalc} 
to Eq.\ref{e:mud} to obtain
\begin{equation}
\tan\phi_2 = {-\cos(l-l_\Omega)\over\cot({\pi\over2}-\delta_\mathrm{GP})\cos b-\sin(l-l_\Omega)\sin b}
\label{e:phi2}\end{equation}
Applied to the same source, $\phi=-\phi_2$, and thus either angle may
be computed with Eq.\ref{e:phi} or with Eq.\ref{e:phi2}.

\section{Proper motions and velocity corrections\label{s:correct}}

\begin{figure}
\centerline{\includegraphics[angle=0,width=0.8\columnwidth]{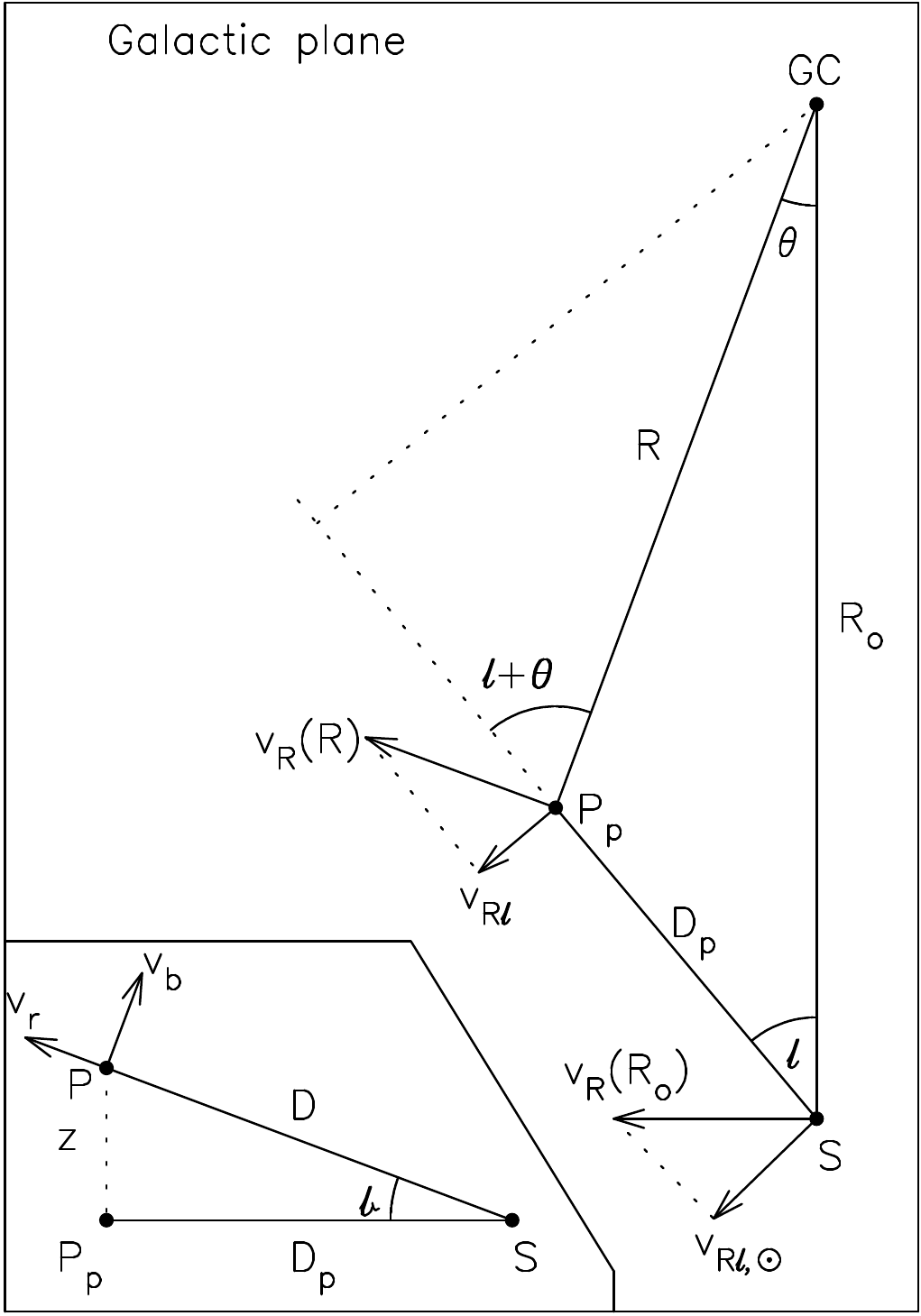}}
\caption{Definition of angles and distances in the galactic plane
($z=0$), and (inset) of the projected distance $D_p$ to the pulsar.
S is the Sun, GC the galactic centre, P the pulsar and P$_{\mathrm{p}}$
the projection of the pulsar position on the galactic plane.
\label{f:syscoor}}
\end{figure}

The space velocity of a star in the Galaxy may be decomposed
into the average space velocity of its surroundings and its velocity
with respect to this average, i.e.\ its peculiar velocity.
The velocity of the local standard of rest for the Sun is its
galactic rotation velocity, $v_R(R_o)$, where $R_o$ is the distance
to the galactic centre.
The peculiar velocity of the Sun is usually written [$U$,$V$,$W$],
where the components are respectively in the direction from the Sun
towards the galactic centre, in the direction of the galactic
rotation, and perpendicular to the galactic plane.
The total velocity of the Sun may thus be written
\begin{equation}
\vec{v}_\odot = [U,V+v_R(R_o),W]
\label{e:vsun}\end{equation}
For a pulsar in the galactic plane, with $b=0$, the velocity of the
local standard of rest is also given by the rotation velocity $v_R(R)$
around the centre of the galaxy, at the galactocentric
distance of the pulsar $R$ (see Fig.\,\ref{f:syscoor}).
This velocity is in the plane of the galaxy, in the direction
perpendicular to the line connection the pulsar to the galactic
center.
For a pulsar far from the plane, the meaning of the Local Standard of
Rest is less obvious, because the halo stars do not participate in the
rotation of the disk. The birthplace of the neutron star is (with the
few exceptions mentioned above) in the galactic plane, therefore we
use for its local standard of rest the galactic rotation
$\vec{v}_\mathrm{R}(R)$ of its projection on the galactic plane.
The total velocity of a pulsar at distance $D$ and
galactic coordinates $l,b$, may be written in the same
coordinate frame as used for the Sun (see Figure\,\ref{f:syscoor}):
\begin{equation}
\vec{v}_p = [U_p+v_R(R)\sin\theta,V_p+v_R(R)\cos\theta,W_p]
\label{e:vpulsar}\end{equation}
with $[U_p,V_p,W_p]$ the peculiar velocity of the pulsar.
To obtain the velocity in the $l$-direction, we write the unit
vector in this direction as:
\begin{equation}
\vec{l} = [-\sin l,\cos l,0]
\label{e:ell}\end{equation}
Thus the observed relative velocity in the longitude direction is
\begin{equation}
\vec{l}\cdot(\vec{v}_p-\vec{v}_\odot) = (\mu_{l*,v}+\mu_{l*,G}) D
\equiv \mu_{l*}D
\label{e:vlobs}\end{equation}
where the peculiar velocity in the longitude direction is
\begin{equation}
 v_l\equiv \mu_{l*}D \equiv -U_p\sin l+V_p\cos l
\label{e:vl}\end{equation}
and the correction for galactic rotation and solar peculiar velocity
is
\begin{equation}
\mu_{l*, G}\,D\equiv U\sin l-[V+v_R(R_o)]\cos l
+v_R(R)\cos(\theta+l)
\label{e:deltamul}\end{equation}
The angle $(\theta+l)$ may be computed from
(see Fig.\,\ref{f:syscoor}):
\begin{equation}
\tan(\theta+l) = {R_o\sin l\over R_o\cos l-D_p} 
= {R_o\sin l\over R_o\cos l-D\cos b} 
\label{e:thetalb}\end{equation}
with $D_p$ the projected distance towards the pulsar.
Eq.\,\ref{e:vsubl} follows from Eqs.\,\ref{e:vlobs}-\ref{e:thetalb}.

The unit vector in the $b$-direction may be written
\begin{equation}
\vec{b} = [-\sin b \cos l,-\sin b\sin l, \cos b]
\end{equation}
and the relative velocity in this direction
\begin{equation}
\vec{b}\cdot(\vec{v}_p-\vec{v}_\odot) = (\mu_{b,v}+\mu_{b, G}) D
\equiv \mu_{b}D
\label{e:vb}\end{equation}
with
\begin{equation}
v_b\equiv \mu_{b,G}\,D = -U_p\sin b\cos l-V_p\sin b\sin l+W_p\cos b
\end{equation}
and
\begin{eqnarray}
\mu_{b, G}D&=&U\sin b\cos l+[V+v_R(R_o)]\sin b\sin l-W\cos b \nonumber \\
& & \label{e:deltamub} -v_R(R)\sin(\theta+l)\sin b
\end{eqnarray}
For a pulsar in direction $l,b$, we can compute $\mu_{l*, G}$ and
$\mu_{b, G}$, as a function of distance $D$ from
Eqs.\ref{e:deltamul}, \ref{e:thetalb} and \ref{e:deltamub}.
Because the rotation of the sum of two vectors is equal to
the sum of two rotated vectors, symbolically: 
$\mathcal{R}(\vec{a}+\vec{\Delta a})=\mathcal{R}(\vec{a})
+\mathcal{R}(\Delta\vec{a})$, we may rotate the corrections
with Eqs.\ref{e:mul}, \ref{e:mub}.
Hence:
\begin{eqnarray}
\mu_{\alpha*, G}&=&\phantom{-}\mu_{l*, G}\cos\phi_2 +
\mu_{b, G}\sin\phi_2  \label{e:rotcora}\\
\mu_{\delta, G} & = & -\mu_{l*, G}\sin\phi_2  + \mu_{b, G}\cos\phi_2\label{e:rotcorb}
\end{eqnarray}
where $\phi_2$ is given by Eq.\ref{e:phi2}.

\section{Numerical evalution of the likelihood in for a
  semi-anisotropic maxwellian}
\label{a:num_int}

To integrate Eq.\,\ref{e:jointmaxwelsi}, we first separate the terms involving the velocity and define
\begin{eqnarray}
I_v &=& e^{-A_3}\int_0^\infty v^2e^{-A_1v^2-A_2v}dv \quad\mathrm{where}
\nonumber \\
 & A_1 &\equiv {1\over2\sigma^2} +
 {1\over2}\left({\sin\xi_1\cos\xi_2\over D\sigma_\alpha}\right)^2+
{1\over2}\left({\sin\xi_1\sin\xi_2\over
    D\sigma_\delta}\right)^2\nonumber \\
 &A_2&\equiv  {\sin\xi_1\over  D}\left[
{(\mu_{\alpha*,G}-\mu'_{\alpha*})\cos\xi_2\over{\sigma_\alpha}^2}+
{(\mu_{\delta,G}-\mu'_\delta)\sin\xi_2\over{\sigma_\delta}^2}\right]\nonumber\\
 & A_3&\equiv {(\mu_{\alpha*,G}-\mu'_{\alpha*})^2\over2{\sigma_\alpha}^2}
 + {(\mu_{\delta,G}-\mu'_\delta)^2\over2{\sigma_\delta}^2}
\end{eqnarray}
The result of this integral is
\begin{equation}
I_v={A_2e^{-A_3}\over4{A_1}^2}\left[\sqrt{\pi}e^{E^2}\left({1\over2E}+E\right)\mathrm{erfc}(E)-1\right];\quad
E\equiv{A_2\over2\sqrt{A_1}}
\end{equation}
Entering this in Eq.\,\ref{e:jointmaxwelsi}, we obtain:
\begin{eqnarray}
L_\mathrm{sim}(\sigma) &=&
\int_o^{D_\mathrm{max}} \int_0^{2\pi} \int_0^\pi
\mathcal{C}f_D(D)
\exp\left[-\,{(1/D-\varpi')^2\over2{\sigma_\varpi}^2}\right]\nonumber \\
& \times&
\sin\xi_1\,2\sqrt{\frac{2}{\pi}}{1\over\sigma^3}
I_v(D,\xi_1,\xi_2)  d\xi_1 d\xi_2 dD
\label{e:psim}\end{eqnarray}
Returning to Eq.\,\ref{e:definejmsi}, we note that for fixed distance $D$,
velocity $v$ and angle $\xi_1$, $P_\mathrm{sim}$ reaches it maximum
when the arguments of the exponents that include the proper motions
are zero. The value of $\xi_2$ for which this is the case follows from
\begin{equation}
\tan\xi_{2m} ={ \mu'_{\alpha*}-\mu_{\alpha*.G}(D)\over\mu'_\delta-\mu_{\delta,G}(D)}
\label{e:ksi2}\end{equation} 
Because this angle is the same for every $v$, the same value of
$\xi_2$ also maximizes the integrand of Eq.\,\ref{e:psim}. The integration of
Eq.\,\ref{e:psim} is done in three steps.
First we fix $D$ and $\xi_1$, and determine the range of
$\xi_2$ from the condition Eq.\,\ref{e:condition} (or equivalently
by testing with Eq.\,\ref{e:vzcorrect} that $v_z$ is in the
right direction). We divide this range in three parts,
one given by $(\xi_{2m}-h)$ to $(\xi_{2m}+h)$, and the other two
dividing the remaining range, and integrate over $\xi_2$ in
each part separately with a 64-node gaussian quadrature.
We find that $h=2\pi/70$ leads to accurate results.
Second, we integrate over $\xi_1$ with one 64-node gaussian quadrature.
Finally, we integrate over $D$, in steps of 100\,pc, for
$D_\mathrm{max}=10$\,kpc.

We compute $L_\mathrm{maxw-si} (\sigma)$ on a grid of values of $\sigma$,
in steps of 5\,km/s, interpolate linearly to get a grid with steps
of 1\,km/s.

\section{Master list\label{s:master}}

\begin{table*}
\caption{Master list of the pulsars used in our study. \label{t:master} }
\centerline{
\begin{tabular}{cllcr@{\hspace{0.2cm}}rr@{\hspace{0.3cm}}rr@{\hspace{0.3cm}}rr@{\hspace{0.3cm}}rrc}
 & B-name & J-name & & l\phantom{42}  & b\phantom{42}  &  $\varpi'$ & $\sigma_\varpi$
 & $\mu'_{\alpha*}$ & $\sigma_\alpha$ & $\mu'_\delta$ & $\sigma_\delta$
 & $\tau_\mathrm{c}$\phantom{m} & ref \\
       &  & & & ($^\circ$)\phantom{4} & ($^\circ$)\phantom{4} & \multicolumn{2}{c}{(mas)} &
         \multicolumn{2}{c}{\phantom{m}(mas/yr)} & \multicolumn{2}{c}{\phantom{mm}(mas/yr)} & (Myr) \\
\hline
 1 & & J0034$-$0721 &   & 110.42 & $-$69.82 &   0.93 &   0.08 &  10.37 &   0.08 &  $-$11.13 &   0.16 &  36.7 & 6  \\
 2 & & J0108$-$1431 &   & 140.93 &$-$76.82 &   4.17 &   1.42 &  75.05 &   2.26 & $-$152.54 &   1.65 & 166.4 & 5  \\
 3 & B0136+57 & J0139+5814 &   & 129.22 &  $-$4.04 &   0.37 &   0.04 & $-$19.11 &   0.07 &  $-$16.60 &   0.07 &   0.4 & 6  \\
 4 & B0329+54 & J0332+5434 & i & 145.00 &  $-$1.22 &   0.94 &   0.11 &  17.00 &   0.27 &   $-$9.48 &   0.37 &   5.5 & 1  \\
 5 & B0355+54 & J0358+5413 & i & 148.19 &   0.81 &   0.91 &   0.16 &   9.20 &   0.18 &    8.17 &   0.39 &   0.6 & 4  \\
\\
 6 & B0450+55 & J0454+5543 &   & 152.62 &   7.55 &   0.84 &   0.05 &  53.34 &   0.06 &  $-$17.56 &   0.14 &   2.3 & 6  \\
 7 & & J0538+2817 & i &179.72 &  $-$1.69 &   0.72 &   0.12 & $-$23.57 &   0.10 &   52.87 &   0.10 &   0.6 & 6  \\
 8 & B0628-28 & J0630$-$2834 &    &236.95 & $-$16.76 &   3.01 &   0.41 & $-$46.30 &   0.99 &   21.26 &   0.52 &   2.8 & 5  \\
 9 & B0656+14 & J0659+1414 & i & 201.11 &   8.26 &   3.47 &   0.36 &  44.07 &   0.63 &   $-$2.40 &   0.29 &   0.1 & 2  \\
10 & B0809+74  & J0814+7429 &   & 140.00 &  31.62 &   2.31 &   0.04 &  24.02 &   0.09 &  $-$43.96 &   0.35 & 122.0 & 1  \\
\\
11 & B0818-13  & J0820$-$1350 &   & 235.89 &  12.59 &   0.51 &   0.04 &  21.64 &   0.09 &  $-$39.44 &   0.05 &   9.3 & 6  \\
12 & B0919+06 & J0922+0638 &   & 225.42 &  36.39 &   0.83 &   0.13 &  18.35 &   0.06 &   86.56 &   0.12 &   0.5 & 3  \\
13 & B0950+08 & J0953+0755 &   & 228.91 &  43.70 &   3.82 &   0.07 &  $-$2.09 &   0.08 &   29.46 &   0.07 &  17.5 & 1  \\
14 & B1133+16 & J1136+1551 &   & 241.90 &  69.20 &   2.80 &   0.16 & $-$73.95 &   0.38 &  368.05 &   0.28 &   5.0 & 1  \\
15 & B1237+25 & J1239+2453 &   & 252.45 &  86.54 &   1.16 &   0.08 &$-$106.82 &   0.17 &   49.92 &   0.18 &  22.9 & 1  \\
\\
16 & B1508+55 & J1509+5531 &   & 91.33 &  52.29 &   0.47 &   0.03 & $-$73.64 &   0.05 &  $-$62.65 &   0.09 &   2.3 & 6  \\
17 & B1541+09 & J1543+0929 &   & 17.81 &  45.78 &   0.13 &   0.02 &  $-$7.61 &   0.06 &   $-$2.87 &   0.07 &  27.5 & 6  \\
18 & B1556-44 & J1559$-$4438 &   & 334.54 &   6.37 &   0.38 &   0.08 &   1.52 &   0.14 &   13.15 &   0.05 &   4.0 & 5  \\
19 & B1929+10 & J1932+1059 & i & 47.38 &  $-$3.88 &   2.78 &   0.06 &  94.06 &   0.09 &   43.24 &   0.17 &   3.1 & 7  \\
20 & & J1935+1616 &   & 52.44 & $-$2.09 &   0.22 &   0.12 &   1.13 &   0.13 &  $-$16.09 &   0.15 &   0.9 & 6  \\
\\
21 & B2016+28 & J2018+2839 &   &  68.10 &  $-$3.98 &   1.03 &   0.10 &  $-$2.64 &   0.21 &   $-$6.17 &   0.38 &  59.8 & 1  \\
22 & B2020+28 & J2022+2854 &   &  68.86 &  $-$4.67 &   0.61 &   0.08 &  $-$3.46 &   0.17 &  $-$23.73 &   0.21 &   2.9 & 7  \\
23 & B2021+51 & J2022+5154 &   &  87.86 &   8.38 &   0.78 &   0.07 &  $-$5.03 &   0.27 &   10.96 &   0.17 &   2.7 & 7  \\
24 & B2045-16 & J2048$-$1616 &   &  30.51 & $-$33.08 &   1.05 &   0.03 & 113.16 &   0.02 &   $-$4.60 &   0.28 &   2.8 & 6  \\
25 & B2053+36 & J2055+3630 &   & 79.13 &  $-$5.59 &   0.17 &   0.03 &   1.04 &   0.04 &   $-$2.46 &   0.13 &   9.5 & 6  \\
\\
26 & & J2144$-$3933 & i &  2.79 & $-$49.47 &   6.05 &   0.56 & $-$57.89 &   0.88 & $-$155.90 &   0.54 & 272.3 & 5  \\
27 & B2154+40 & J2157+4017 &   & 90.49 & $-$11.34 &   0.28 &   0.06 &  16.13 &   0.10 &    4.12 &   0.12 &   7.1 & 6  \\
28 & B2310+42 & J2313+4253 &   & 104.41 & $-$16.42 &   0.93 &   0.07 &  24.15 &   0.10 &    5.95 &   0.13 &  49.3 & 6  \\
\end{tabular}
}
\tablefoot{The last column
  gives the reference in Table\,\ref{t:sources} from which the parallax with
  error (columns 7,8), and the proper motions with their errors
  (columns 9-12) are taken. In the case of asymmetric errors we take
  the larger one. Columns 11 gives the characteristic age   $\tau_c\equiv P/(2\dot P)$.
  An i in column 5  indicates that the model velocity distribution for this pulsar is
  isotropic in the models that mix isotropic and semi-isotropic velocity distributions.}
\end{table*}

\end{appendix}
\end{document}